\documentclass[%
 reprint,
nobibnotes,
bibnotes,
 amsmath,amssymb,
 aps,
pre,
]{revtex4-1}
\usepackage{natbib} 
\usepackage{xcolor}

\usepackage{soul}
\usepackage{hyperref}
\usepackage{wasysym} 
\usepackage{graphicx}
\usepackage{tikz}
\usetikzlibrary{calc}
 \usetikzlibrary{arrows.meta}
\usepackage{dcolumn}
\usepackage{bm}

\tikzstyle{MyVertex}=[draw,circle,fill=black,inner sep=0,minimum size=3pt]
\tikzstyle{MyVacantVertex}=[draw,circle,fill=white,inner sep=0,minimum size=4pt]

\tikzstyle{MyDashedEdge}=[draw,dotted,line width=1pt]
\tikzstyle{MyFullEdge}=[line width=1pt]

\begin{document}

\preprint{Draft}

\title{Exact solutions for models of evolving clustered networks with addition and deletion via age-resolved message passing}

\author{Peter Mann}
\email{peter.mann@diai.io}
\affiliation{Data Insights AI, CodeBase Edinburgh, Argyle House, 3 Lady Lawson Street, Edinburgh, EH3 9DR, United Kingdom}

\author{Lei Fang}
\affiliation{School of Computer Science, University of St Andrews, St Andrews, Fife KY16 9SX, United Kingdom }

\author{Simon Dobson}
\affiliation{School of Computer Science, University of St Andrews, St Andrews, Fife KY16 9SX, United Kingdom }

\date{\today}

\begin{abstract}
Evolving networks experience vertex addition and deletion over time via processes which shape their structure and function. In many empirical networks, vertices do not arrive or depart in isolation but instead within groups or cohorts: for example, coauthorship networks grow by the addition of cliques whenever a new paper with multiple authors is published. In this paper, we propose a stochastic network evolution model based on block graphs that evolve through the addition of fully connected subgraphs ($m$-cliques) and the deletion of individual vertices. We derive a master equation governing the time evolution of the joint degree distribution and obtain an exact closed-form solution, valid for an arbitrary vertex deletion rate under uniform attachment, that describes both growing and constant-size networks. From this solution we obtain the marginal degree distributions and the clustering coefficient in closed form and the size of the giant connected component and percolation threshold governing robustness to random vertex failure using age-resolved message passing. Our results generalise existing single-vertex addition-deletion models to clique-based dynamics at arbitrary rates of turnover, and are confirmed by Monte Carlo simulation.
\end{abstract}

\pacs{Valid PACS appear here}
\maketitle


\section{Introduction}
\label{sec:introduction}

Networks provide a powerful mathematical abstraction for representing relationships among interacting entities, with applications spanning the natural and social sciences, engineering, and commerce. The study of networks has emerged as a unifying framework across disciplines, from the analysis of protein-protein interactions in biology to the mapping of social structures in sociology~\cite{Newman_2019, Newman_Strogatz_Watts_2001}. Network science draws upon graph theory, statistical mechanics, and computer science to develop predictive models of complex phenomena, offering insights into the structural and dynamical properties that govern real-world systems.

A fundamental challenge in network science is understanding how networks evolve over time \cite{Moore_Ghoshal_Newman_2006,Ghoshal_Chi_Barabasi_2013,Budnick_Biham_Katzav_2025,Kong_Sarshar_Roychowdhury_2008,Miura_Takayasu_Takayasu_2012,Bagrow_Brockmann_2013,Fasino_Tonetto_Tudisco_2020,Hartle_Papadopoulos_Krioukov_2021,Ke_Yi_2004,Ben-Naim_Krapivsky_2007,Christensen_Thakar_Albert_2007,Deijfen_Lindholm_2009,Zhang_He_Rayman-Bacchus_2016,Bauke_Moore_Rouquier_Sherrington_2011,Ghoshal_Newman_2007,Ambrosio_Cerulli_Serra_Sorgente_Vaccaro_2025,Giroire_Nisse_Ohulchanskyi_Sulkowska_Trolliet_2023}. Real-world networks are rarely static; they grow through the addition of new vertices and edges, and they contract through the removal of existing ones. This dynamical perspective has motivated the development of evolving network models that capture the time-dependent behaviour of complex systems~\cite{Moore_Ghoshal_Newman_2006}. Early work on network growth, including the celebrated preferential attachment mechanism, demonstrated that simple local rules can generate scale-free degree distributions observed empirically in the World Wide Web, citation networks, and biological systems~\cite{Ghoshal_Chi_Barabasi_2013,Kong_Sarshar_Roychowdhury_2008}. More recent efforts have extended these models to incorporate both addition and deletion of vertices, recognising that many networks experience turnover as well as growth~\cite{Moore_Ghoshal_Newman_2006, Miura_Takayasu_Takayasu_2012,Bagrow_Brockmann_2013,Budnick_Biham_Katzav_2025}. Such addition-deletion models have proven essential for understanding systems ranging from social networks, where users join and depart, to infrastructure networks subject to component failures and replacements.

A distinctive feature of many empirical networks is a high degree of clustering, the tendency for two neighbours of a vertex to also be neighbours of one another \cite{Newman_Strogatz_Watts_2001,Karrer_Newman_2010,Mann_Smith_Mitchell_Dobson_2021,PhysRevLett.103.058701,Mann_Smith_Mitchell_Jefferson_Dobson_2021,PhysRevE.105.044314,Mann_Fang_Dobson_2025,Miller_2009,PhysRevE.80.020901,Hasegawa_Mizutaka_2020,Gleeson_2009}. This property, quantified by the clustering coefficient, is markedly higher in real networks than would be expected from random graph models~\cite{Watts_Strogatz_1998, Newman_2003}. Clustering reflects the presence of tightly knit communities and cohesive substructures within larger networks. To capture this phenomenon, researchers have developed random clustered graph models that explicitly incorporate triangles and higher-order motifs into the generative process~\cite{Newman_Strogatz_Watts_2001,Karrer_Newman_2010,Miller_2009}. These models have enabled analytical treatment of epidemic spreading, percolation thresholds, and other dynamical processes on clustered networks~\cite{Gleeson_2009,Hasegawa_Mizutaka_2020,Mann_Smith_Mitchell_Dobson_2021,Mann_Smith_Mitchell_Jefferson_Dobson_2021,PhysRevE.105.044314,Mann_Fang_Dobson_2025}. However, existing random clustered graph models are predominantly static, describing equilibrium ensembles without explicit consideration of the temporal processes through which clustered structures emerge or dissolve.

A notable gap in the literature is the absence of addition-deletion models for networks in which vertices arrive and depart within cohorts or groups rather than in isolation \cite{PhysRevE.65.026107, PhysRevLett.85.4633, PhysRevE.67.056104}. In many real-world systems, clustering arises naturally because entities join the network as part of a fully connected subgraph, a clique. Coauthorship networks provide a canonical example: when a scientific paper is published, all authors become connected to one another, forming a complete subgraph whose size equals the number of co-authors~\cite{Newman_2004}. Similar structures arise in social networks, where groups of friends form simultaneously. The evolution of such networks cannot be adequately described by models in which vertices arrive singly, as the inherent group structure fundamentally shapes the resulting topology.

Supply chain networks represent another domain where clustered, dynamically evolving structures are of paramount importance. Modern supply chains are not simple linear sequences of suppliers and buyers but rather complex webs of interdependent firms connected through multi-party contractual relationships~\cite{Perera_2017}. A supplier may serve multiple manufacturers simultaneously, while a manufacturer may source components from overlapping sets of suppliers, creating clique-like structures of mutual dependency. These networks are highly dynamic: new partnerships form as firms expand operations, and existing relationships dissolve due to market forces, disruptions, or strategic realignment. Understanding the structural evolution of supply chain networks is critical for assessing resilience, the ability to withstand and recover from disruptions such as natural disasters, geopolitical conflicts, or pandemic-induced shocks~\cite{Wagner_Neshat_2010}. Network modelling approaches based on graph theory have been employed to analyse the topological properties that confer robustness, including redundancy, connectivity, and the distribution of critical nodes. However, analytical models that can capture the time evolution of supply chain structure under realistic addition-deletion dynamics remain underdeveloped.

The structure of data storage and retrieval systems provides a further motivation for studying dynamically evolving clustered graphs.
Graph databases have emerged as a powerful paradigm for managing interconnected data, finding applications in knowledge graphs, social platforms, and enterprise systems.
These databases can exhibit growth (and contraction) based on structured addition of data.
As artificial intelligence systems increasingly rely on graph-structured data for tasks such as recommendation, fraud detection, and knowledge reasoning, understanding the topological properties of these structures becomes essential for computational efficiency.
The energy consumption of data centres powering AI systems has become a significant concern, with graph-based architectures offering more energy-efficient alternatives to attention-based models for certain classes of structured data problems.
Accurate models of graph databases' dynamic structure could inform the design of more efficient storage layouts, indexing strategies, and query optimisation algorithms, contributing to the broader goal of sustainable computing.

In this paper, we introduce an addition-deletion process in which vertices arrive and depart within fully connected groups, or cliques, rather than individually.
Our model is based on evolving block graphs in which every biconnected component is a clique, and captures the essential features of networks that grow through cohort addition while experiencing turnover.
We derive a general rate equation governing the time evolution of the joint degree distribution, where the joint degree of a vertex encodes its participation in cliques of different sizes.
For an arbitrary rate of vertex deletion, encompassing both growing networks and constant-size networks in which addition and deletion are balanced, with uniform attachment kernels, we obtain an exact analytical solution for the asymptotic joint degree distribution using generating function techniques and the method of characteristics.
Our results generalise the single-vertex addition-deletion model of Moore \textit{et al.}~\cite{Moore_Ghoshal_Newman_2006} to the case of clique-based dynamics.
We derive closed-form expressions for the marginal degree distributions and the clustering coefficient, and, through an age-resolved message passing, obtain the size of the giant connected component and the percolation threshold that governs the network's robustness to random vertex failure.
Monte Carlo simulations confirm the accuracy of our analytical predictions.
These results provide a foundation for understanding the structure and dynamics of non-locally-treelike networks subject to realistic turnover processes.

\section{2,3-clique Model}
\label{sec:model1}

We begin the exposition by treating a simple scenario of 2- and 3-cliques to examine the addition-deletion mechanism in detail. We will generalise the mechanics to arbitrary clique sizes in the following section \cite{PhysRevLett.103.058701}.

The joint degree tuple $(s,t)$ indicates a vertex's membership in $s$ ordinary edges and $t$ triangles, such that the overall degree is $k=s+2t$. Let the joint degree distribution $p_{s,t}$ of our network be the fraction of vertices connected to $s$ single edges and $t$ triangles. As with any probability distribution we have the condition
\begin{equation}
    \sum_s\sum_t p_{s,t}=1\label{eq:pst_norm}.
\end{equation}
The average $s$ and $t$ degrees of a randomly selected vertex are given by
\begin{equation}
    \langle s\rangle = \sum_s\sum_tsp_{s,t}, \qquad   \langle t\rangle = \sum_s\sum_ttp_{s,t}.
\end{equation}
A given edge belongs to at most one clique, \textit{i.e.}, the cliques are edge-disjoint; however, a vertex can belong to many cliques.

Consider a stochastic process which evolves in discrete time steps. At a given time step, a clique of size $m\in \{2,3\}$ is added to the network. There are $m$ vertices in the clique and we constrain their joint degrees to be equal to $(s',t')$. There are the necessary consistency conditions on the values of $s'$ and $t'$ such that $s'$ cannot be zero if $m=2$ and likewise $t'\neq 0$ if $m=3$. Vertices in $m=2$-cliques have a spare $(s'-1,t')$ stubs to connect, whilst vertices in $m=3$-cliques form $s'$ ordinary edges and $t'-1$ new triangles.

Concretely, the addition part of a time step proceeds as follows. (i) \emph{Cohort creation:} the $m$ new vertices are created and joined pairwise, forming the incoming $m$-clique; this clique supplies one of the prescribed memberships of each new vertex (one of its $s'$ edge memberships if $m=2$, or one of its $t'$ triangle memberships if $m=3$). (ii) \emph{Membership completion:} each new vertex then independently realises its remaining memberships: for each remaining 2-clique membership it selects one existing vertex at random, with probabilities set by the attachment kernels introduced below, and places a single edge to it, and for each remaining 3-clique membership it selects \emph{two} existing vertices in the same manner and places all three edges amongst itself and the selected pair. Every new triangle therefore contains exactly one incoming vertex and two existing vertices, each of which gains one triangle membership. For example, when a 3-clique is added with $t'=2$, each of the three new vertices forms $t'-1=1$ additional triangle in this manner.

Let $\pi_{2,(s,t)}p_{s,t}$ be the probability that a 2-clique edge connects to an existing vertex of joint degree $(s,t)$. Let $\pi_{3,(s,t)}p_{s,t}$ be the probability that a triangle stub connects in one corner to an existing vertex with joint degree $(s,t)$. There is an attachment kernel for each kind of incident stub we must pair up. Given that the vertices must connect somewhere in the network we have the following normalisation conditions
\begin{equation}
    \sum_s\sum_t \pi_{2,(s,t)}p_{s,t} = 1, \qquad \sum_s\sum_t \pi_{3,(s,t)}p_{s,t} = 1.
\end{equation}
These topology-dependent functions generalise the attachment kernels used in previous models of network growth for our purpose \cite{Krapivsky_Redner_2001, Moore_Ghoshal_Newman_2006}. Under the uniform attachment analysed in Sec.~\ref{sec:uniform}, these selections are uniformly random. Other choices of kernels, such as preferential attachment based on degree or given joint degrees have been shown to generate power law networks \cite{Moore_Ghoshal_Newman_2006}, we leave this for future study.

\begin{figure*}[t]
  \centering
  \includegraphics[width=0.95\textwidth]{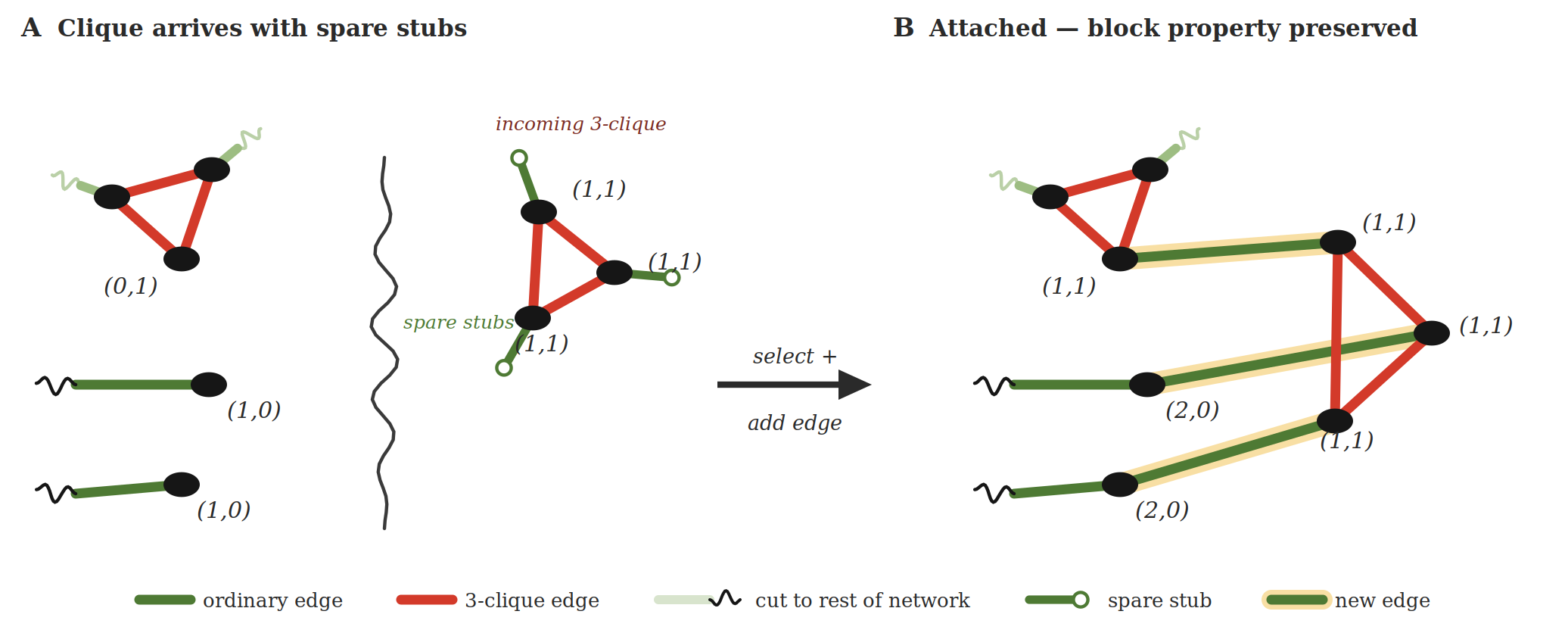}
\caption{The clique-addition step of the model. An incoming $m$-clique (here a $3$-clique, red) arrives carrying a fixed joint degree, a prescribed set of spare stubs on its vertices (open circles). Each spare stub of the incoming vertices is resolved by selecting an existing vertex at random (via the attachment kernel) and adding a new edge incrementing the selected vertex's joint degree. A triangle stub would recruit two existing vertices and add all three edges among the triple. The receiving vertices connect to the wider network only across the cut boundaries (accidental closure vanishes for large networks), so every biconnected component remains a clique.}
  \label{fig:clique-addition}
\end{figure*}

We assume that the graph is sufficiently large such that the effects of accidental clique formation or a given edge belonging to multiple cliques is negligible. Thus, the resulting graph is edge-disjoint and in the limit of large $n$, every biconnected component (block) is a clique, see Fig \ref{fig:clique-addition}. 
Within the same time step we will also allow for the removal of up to $m$ vertices with the probability of removal being $r$. To perform the removals we draw $m$ vertices from the graph at random and delete them (and all of their edges) independently with probability $r$.

\subsection{Rate Equation for $p_{s,t}$}
\label{sec:model1:sub:RE}

At a given time step $\tau$ let the probability of selecting a vertex at random from the graph with joint degree $s,t$ be $p_{s,t}(\tau)$. The number of vertices with this joint degree is $np_{s,t}(\tau)$. In the next time step the number is $(n+m-rm)p'_{s,t}(\tau+1)$, where $p'_{s,t}(\tau+1)$ is the new value of $p_{s,t}(\tau)$ at $\tau+1$. The evolution of the joint degree distribution is governed by a rate equation as follows. 

We begin with the number of vertices that had joint degree $s,t$ in the previous time step $np_{s,t}(\tau)$. There is the direct contribution $m\delta_{s,s'}\delta_{t,t'}$ of the new vertices if $s$ and $t$ happen to equal $s'$ and $t'$, where $\delta_{i,j}$ is the Kronecker delta
\begin{equation}
    \delta_{i,j} = 
    \begin{cases}
        1,\qquad i=j,\\
        0,\qquad i\neq j
    \end{cases}
\end{equation}
Considering $m=2$, there are two vertices in the new clique that each need to form $s'-1$ new ordinary edges and $t'$ new triangles. The change in the joint degree distribution as a result of the new ordinary edge connections is
\begin{equation}
    2(s'-1)\left(\pi_{2,(s-1,t)}p_{s-1,t} - \pi_{2,(s,t)}p_{s,t}\right),
\end{equation}
where the leading $2$ accounts for both vertices in the new clique and $s'-1$ is the number of new 2-clique edges being created per vertex. New $s$-edge connections to vertices with previous joint degree $(s-1,t)$ leads to a probability mass increase in $p_{s,t}$, whilst connections to vertices that started with joint degree $(s,t)$ lead to loss of probability mass. 

Each vertex in the 2-clique forms $t'$ new triangles to existing vertices in the network. This leads to the following terms in the rate expression
\begin{equation}
    2\cdot 2 \cdot t'\left(\pi_{3,(s,t-1)}p_{s,t-1} - \pi_{3,(s,t)}p_{s,t}\right).
\end{equation}
The first term tracks the flux into joint degree $(s,t)$ due to existing vertices with joint degree $(s,t-1)$ forming new triangles; whilst the second term accounts for flux out due to existing vertices with joint degree $(s,t)$ forming new triangles. The leading multiplicative factors are due to there being 2 vertices in the incident 2-clique, each with $t'$ triangles that in turn connect 2 $t$-edge stubs per triangle.

Moving now to the second scenario we set $m=3$ and note that the 3 new vertices each with joint degree $(s',t')$ need to create $s'$ new 2-clique edges and $2(t'-1)$ new 3-clique edges or equivalently $t'-1$ triangles to existing vertices. The contribution to the rate equation is
\begin{equation}
    3\cdot 2\cdot  (t'-1)\cdot \left(\pi_{3,(s,t-1)}p_{s,t-1} - \pi_{3,(s,t)}p_{s,t}\right),
\end{equation}
where the leading multiplicative factor accounts for 3 vertices each creating $t'-1$ new triangles, each with 2 3-clique edges per triangle. The first term in the bracket accounts for the flux into $p_{s,t}$ from vertices that had joint degree $(s,t-1)$; the second accounts for the loss of density due to vertices that originally had joint degree $(s,t)$ flowing to $(s,t+1)$. Additionally, the new 3-clique forms $s'$ new 2-clique edges to existing vertices. We see net migration into $p_{s,t}$ from vertices that originally had joint degree $(s-1,t)\rightarrow (s,t)$; and flux out to $(s,t)\rightarrow (s+1,t)$
\begin{equation}
    3\cdot s'\cdot\left(\pi_{2,(s-1,t)}p_{s-1,t} - \pi_{2,(s,t)}p_{s,t}\right).
\end{equation}

We now turn to the deletion part of the process. To remove vertices from the network we select $m$ vertices uniformly at random from the graph, removing them and all of their connections. 
When a neighbour of a vertex $v$ is deleted, the topology of the clique shared by $v$ and the deleted neighbour is degraded. If the connection was an ordinary edge (2-clique), the edge is simply removed.
However, if the connection was a triangle (3-clique), the triangle degrades into an ordinary edge (because removing the vertex removes \textit{two} edges from the triangle, leaving the third untouched).

The positive flux \textit{into} state $(s,t)$ arises from two sources: (i) neighbours connected via 2-clique edges are deleted, causing the vertex to transition from $(s+1, t) \to (s,t)$; and (ii) neighbours connected via triangles are deleted, breaking the triangle. This reduces $t$ by 1 but leaves an ordinary edge, increasing $s$ by 1. In this case, the vertex transitions from $(s-1, t+1) \to (s,t)$.

The contributions are
\begin{equation}
    rm(s+1)p_{s+1,t} + 2 rm (t+1)p_{s-1,t+1}.
\end{equation}
The factor of 2 in the second term accounts for the fact that a triangle has two other vertices; the deletion of \emph{either} neighbour causes the triangle to degrade.

Similarly, we must account for the flux \textit{out} of state $(s,t)$ due to neighbour deletion. A vertex with $s$ edges loses density to $(s-1, t)$ at a rate proportional to $s$. A vertex with $t$ triangles loses density to $(s+1, t-1)$ at a rate proportional to $2t$. Finally, the vertex itself may be selected for deletion. Combining these loss terms, the total removal and degradation term is
\begin{equation}
    -rm(s + 2t + 1)p_{s,t}.
\end{equation}

With all of the terms that account for the addition-deletion process in place, we can now write the rate equations for the case of $m=2$
\begin{align}
    (n+m-mr)p'_{s,t} =\ & np_{s,t} + m\delta_{s,s'}\delta_{t,t'} \nonumber\\
    &+  2(s'-1)\left(\pi_{2,(s-1,t)}p_{s-1,t} - \pi_{2,(s,t)}p_{s,t}\right) \nonumber\\
    &+  4 t'\left(\pi_{3,(s,t-1)}p_{s,t-1} - \pi_{3,(s,t)}p_{s,t}\right) \nonumber\\
    &+ rm(s+1)p_{s+1,t} + 2rm (t+1)p_{s-1,t+1} \nonumber\\
    &- rm\left(s +2t+1\right)p_{s,t}.\label{eq:rate-2}
\end{align}
When $m=3$ we have a similar expression
\begin{align}
    (n+m-mr)p'_{s,t} =\ & np_{s,t} + m\delta_{s,s'}\delta_{t,t'} \nonumber\\
    &+  3 s' \left(\pi_{2,(s-1,t)}p_{s-1,t} - \pi_{2,(s,t)}p_{s,t}\right) \nonumber\\
    &+6  (t'-1) \left(\pi_{3,(s,t-1)}p_{s,t-1} - \pi_{3,(s,t)}p_{s,t}\right)\nonumber\\
    &+ rm(s+1)p_{s+1,t} + 2rm (t+1)p_{s-1,t+1} \nonumber\\
    &- rm\left(s +2t+1\right)p_{s,t}.\label{eq:rate-3}
\end{align}
Having constructed the rate expressions for the 2- and 3-clique model, we will now turn to the general model containing an arbitrary size and number of cliques.

\section{n-clique model }
\label{sec:model2}

We can now generalise the model to the case of an arbitrary number of clique topologies that can be added or deleted from the network, beyond just 2- and 3-cliques. 

We define the \textit{topology set}, $\bm\chi\subset \mathbb{N}$, to be the set of all clique sizes in the model up to some maximum size. We assume that $\bm\chi$ is consecutive, containing every size from ordinary edges up to the maximum, $\bm\chi=\{2,3,\dots,\bar\alpha_m\}$; isolated vertices (1-cliques) do not belong to the topology set. Let variables with a bar $\bar \alpha_1, \dots, \bar \alpha_m\in \bm \chi$ be clique sizes in the topology set and let their un-barred counterparts $\alpha_1, \dots,\alpha_m$ index the number of cliques of sizes $\bar \alpha_1$, $\bar \alpha_2$ and so on.

Throughout, the subscript $j\in\{1,\dots,m\}$ is a \textit{class label} rather than a clique size: since the set is consecutive, class $j$ has size $\bar\alpha_j=j+1$, so that the smallest class, $j=1$, is that of ordinary edges and $\alpha'_1$ counts the edge memberships of an incoming vertex (the quantity denoted $s'$ in Sec.~\ref{sec:model1}), which need not vanish. Sums and products written over $\bar\alpha_j\in\bm\chi$ are shorthand for sums and products over the class label $j=1,\dots,m$. As in the previous section we use primes $\alpha_1',\dots ,\alpha_m'$ to indicate the fixed joint degree of incoming vertices, as distinct from the joint degrees of vertices already present in the network.

The joint degree of a vertex is now an $m$-tuple that denotes its participation in cliques of given sizes $(\alpha_1,\alpha_2,\dots,\alpha_m)$. The  probability distribution that a randomly selected vertex has a given joint degree is $p_{\alpha_1,\dots,\alpha_m}$ and we have the normalisation
\begin{equation}
\sum_{\alpha_1=0}^\infty\cdots\sum_{\alpha_m=0}^\infty
    p_{\alpha_1,\dots,\alpha_m}=1\label{eq:pstm_norm}
\end{equation}
When adding a clique of size $\bar \alpha_i$, whose vertices each carry the fixed joint degree $(\alpha_1',\dots ,\alpha_m')$ (inclusive of the membership in the incoming $\bar\alpha_i$-clique itself), and removing $\bar\alpha_i$ randomly selected vertices, the correct generalisation of the rate equations in \eqref{eq:rate-2} and \eqref{eq:rate-3} is 
\begin{widetext}

\begin{align}
(n+\bar\alpha_i-r\bar\alpha_i )p'_{\alpha_1,\dots,\alpha_m} =\ & np_{\alpha_1,\dots,\alpha_m} + \bar \alpha_i\prod_{j=1}^{m}\delta_{\alpha_j,\alpha_j'}\nonumber\\
&+ \sum_{\bar\alpha_j\in\bm{\chi}} \lambda_{i,j} \Big( \pi_{\bar\alpha_j,(\alpha_1,\dots,\alpha_j-1,\dots,\alpha_m)}p_{\alpha_1,\dots,\alpha_j-1,\dots,\alpha_m} - \pi_{\bar\alpha_j,(\alpha_1,\dots,\alpha_m)}p_{\alpha_1,\dots,\alpha_m}\Big)\nonumber\\
&+ r\bar\alpha_i \sum_{\substack{\bar\alpha_j\in\bm\chi}} \Big[ (\alpha_j+1)(\bar\alpha_j -1)p_{\alpha_1,\dots,\alpha_{j-1}-1,\alpha_j+1,\dots,\alpha_m} - \alpha_j(\bar\alpha_j -1)p_{\alpha_1,\dots,\alpha_m} \Big]\nonumber\\
&- r\bar\alpha_i\, p_{\alpha_1,\dots,\alpha_m}\label{eq:rate}
\end{align}
where
\begin{equation}
    \lambda_{i,j}=\lambda(\bar\alpha_i,\bar\alpha_j,\alpha_j') = \begin{cases}
\alpha_j' & \text{if } \bar\alpha_i=1 \\
\bar\alpha_i(\bar\alpha_j-1)\Big(\delta_{\bar\alpha_j,\bar \alpha_i}(\alpha_j'-1)
+ (1-\delta_{\bar\alpha_j,\bar\alpha_i})\alpha_j'\Big)  & \text{if } \bar\alpha_i \geq 2
\end{cases}
\label{eq:lambda}
\end{equation}
\end{widetext}
The left hand side of Eq \ref{eq:rate} describes the new state in terms of the updated joint degree distribution $p'_{\alpha_1,\dots,\alpha_m}$ whilst the right hand side is described in terms of the old joint degree distribution  $p_{\alpha_1,\dots,\alpha_m}$ and accounts for the connectivity changes that occur during the time step. From left to right, the first term is the number of vertices that have joint degree $(\alpha_1,\dots,\alpha_m)$ at the start of the time step. The product over the Kronecker delta functions accounts for the probability that $\bar\alpha_i$ new vertices will land in $(\alpha_1,\dots,\alpha_m)$ if the fixed joint degree of the incoming vertices matches this joint degree tuple. 

The next term accounts for the flux of the degrees of existing vertices in the graph as they connect to the incoming clique. It sums the contribution from each clique topology as new edges form in that topology. The prefactor $\lambda_{i,j}=\lambda(\bar\alpha_i,\bar\alpha_j,\alpha_j')$ to this term enumerates the number of new edges that the vertices in the clique will form.

The branch $\bar\alpha_i=1$ of Eq.~\eqref{eq:lambda} describes the addition of a bare vertex forming $\alpha'_1$ ordinary edges only (with $\alpha'_{j>1}=0$); it lies outside the topology set and is retained solely to recover the single-vertex model of Ref.~\cite{Moore_Ghoshal_Newman_2006} in Appendix~\ref{sec:appendix-mapping}. In the clique model proper, $\bar\alpha_i\ge2$. The functions $\pi_{\bar\alpha_j,(\alpha_1,\dots,\alpha_m)}$ are the attachment kernels per topology. When multiplied by $p_{\alpha_1,\dots,\alpha_m}$ they define the probability that an edge within a currently-forming $\bar\alpha_j$-clique connects to a particular pre-existing vertex of joint degree $(\alpha_1,\dots,\alpha_m)$. The attachment kernels obey the following normalisation
\begin{equation}
\sum_{\alpha_1=0}^\infty\cdots\sum_{\alpha_m=0}^\infty\pi_{\bar\alpha_j,(\alpha_1,\dots,\alpha_m)}p_{\alpha_1,\dots,\alpha_m}=1.
\end{equation}
Each incoming vertex carries the fixed joint degree $(\alpha'_1,\dots,\alpha'_m)$ and must therefore create $\alpha_j'$ new cliques of topology $\bar\alpha_j$ (or $\alpha_j'-1$ when $\bar\alpha_j=\bar \alpha_i$, the incoming clique itself supplying one such membership), each clique corner presenting $\bar\alpha_j-1$ edge stubs.

As in Sec.~\ref{sec:model1}, each such membership is realised by the incoming vertex selecting $\bar\alpha_j-1$ existing vertices at random, according to the attachment kernel $\pi_{\bar\alpha_j}$, and adding all edges amongst itself and the selected vertices, completing a new $\bar\alpha_j$-clique in which every selected vertex gains one membership of topology $\bar\alpha_j$. The participation of each vertex in the incoming clique is handled by the Kronecker delta functions. The necessary degree of existing nodes in the network to add a new connection in dimension $\bar\alpha_j$ and result in a joint degree of $(\alpha_1,\dots,\alpha_m)$ is precisely $(\alpha_1,\dots,\alpha_j-1,\dots,\alpha_m)$. 

The remaining two terms account for flux due to the removal process; both due to neighbour-removal and the loss of the removed vertices themselves. Focusing on the first term in square brackets, when a vertex is deleted from a $\bar\alpha_j$-clique, the remaining $(\bar\alpha_j - 1)$ members experience loss of one $\bar\alpha_j$-clique membership and also gain of one $(\bar\alpha_j-1)$-clique membership due to clique degradation. In this case we see a coupled change in the joint degree tuple of the neighbours
\begin{equation}
    (\cdots,\alpha_{j-1}-1,\alpha_{j}+1,\dots)\rightarrow (\cdots,\alpha_{j-1},\alpha_{j},\dots),\label{eq:transitioncoupled}
\end{equation}
where $\bar\alpha_{j-1}$ is the degraded clique topology of $\bar\alpha_j$; and $\alpha_{j-1}$ and $\alpha_{j}$ are the number of those cliques, respectively. The expression assumes that the degradation partner is in $\bm\chi$ and has the convention that 2-cliques simply remove the edge. The left side of Eq \ref{eq:transitioncoupled} is the necessary starting joint degree to both lose a $j$ clique and gain a $j-1$ clique and result in joint degree $\alpha_{j-1}$ and $\alpha_j$ along those two topologies. The leading prefactor counts the number of edges in the clique $(\alpha_j+1)(\bar\alpha_j-1)$ and therefore weights the positive flux into joint degree $(\alpha_1, \dots,\alpha_m)$ by the number of edges per topology. The second subtraction term in the square brackets accounts for loss of probability mass from state $(\alpha_1, \dots,\alpha_m)$ due to removal of neighbours from vertices already in that state. The final $r$-term accounts for the direct removal of probability mass from state $(\alpha_1, \dots,\alpha_m)$ due to removal.

In the asymptotic limit $\tau\rightarrow \infty$ the joint degree distribution reaches a stationary state, the fixed point $p'_{\alpha_1,\dots,\alpha_m}=p_{\alpha_1,\dots,\alpha_m}$, and we find the following expression:

\begin{widetext}

\begin{align}
\label{eq:rate-asymptote-general} 
\bar \alpha_i\prod_{j=1}^{m}\delta_{\alpha_j,\alpha_j'}+&
     \sum_{\bar\alpha_j\in\bm{\chi}} \lambda_{i,j}\Big( \pi_{\bar\alpha_j,(\alpha_1,\dots,\alpha_j-1,\dots,\alpha_m)}p_{\alpha_1,\dots,\alpha_j-1,\dots,\alpha_m} - \pi_{\bar\alpha_j,(\alpha_1,\dots,\alpha_m)}p_{\alpha_1,\dots,\alpha_m}\Big)\nonumber\\
&+r\bar \alpha_i\sum_{\bar\alpha_j\in \bm\chi}(\alpha_j+1)(\bar\alpha_j-1)p_{\alpha_1,\dots,\alpha_{j-1}-1,\alpha_j+1, \dots,\alpha_m} -r\bar \alpha_i \sum_{
\bar\alpha_j\in\bm\chi}\alpha_j(\bar\alpha_j-1)p_{\alpha_1, \dots,\alpha_m} -\bar\alpha_i p_{\alpha_1,\dots,\alpha_m} =0
\end{align}
Let us define the following generating functions
\begin{align}
    g(z) =& \sum_{\alpha_1=0}^\infty \cdots \sum_{\alpha_m=0}^\infty  p_{\alpha_1,\dots,\alpha_m}\prod_{j=1}^mz_{\bar \alpha_j}^{\alpha_j},\label{eq:g}\\
    f_{\bar \alpha_i}(z) =& \sum_{\alpha_1=0}^\infty \cdots \sum_{\alpha_m=0}^\infty  \pi_{\bar\alpha_i,(\alpha_1,\dots,\alpha_m)}p_{\alpha_1,\dots,\alpha_m}\prod_{j=1}^mz_{\bar \alpha_j}^{\alpha_j},
\end{align}
where $z=(z_{\bar \alpha_1}, \dots,z_{\bar \alpha_m})$.
Multiplying Eq \ref{eq:rate-asymptote-general} by $z_{\bar \alpha_1}^{\alpha_1}\cdots z_{\bar \alpha_m}^{\alpha_m}$ and summing over all possible joint degree configurations $(\alpha_1,\dots, \alpha_m)$ we obtain
\begin{align}
\bar{\alpha}_i \prod_{j=1}^{m} z_{\bar{\alpha}_j}^{\alpha'_j} &+ \sum_{\bar{\alpha}_j\in\bm{\chi}} \lambda_{i,j}\Big(z_{\bar{\alpha}_j}-1\Big)f_{\bar{\alpha}_j}(z)+ r\bar{\alpha}_i\sum_{\bar\alpha_j \in \bm{\chi}}\gamma_j(z_{\bar{\alpha}_{j-1}}-z_{\bar{\alpha}_j})\frac{\partial g}{\partial z_{\bar{\alpha}_j}} - \bar{\alpha}_i g(z) = 0\label{eq:main-result}
\end{align}
where we have defined $\gamma_j=(\bar\alpha_j-1)$ and have used the following derivatives
\begin{align}
       \frac{\partial g}{\partial z_{\bar\alpha_j}}&=\sum_{\alpha_1=0}^\infty\cdots\sum_{\alpha_m=0}^\infty(\alpha_j+1)p_{\alpha_1,\dots,\alpha_j+1,\dots,\alpha_m}\prod^m_{k=1}z^{\alpha_k}_{\bar\alpha_k},\\
    z_{\bar\alpha_j} \frac{\partial g}{\partial z_{\bar\alpha_j}}&=\sum_{\alpha_1=0}^\infty\cdots\sum_{\alpha_m=0}^\infty\alpha_jp_{\alpha_1,\dots,\alpha_m}\prod^m_{k=1}z^{\alpha_k}_{\bar\alpha_k},\\
z_{\bar\alpha_{j-1}} \frac{\partial g}{\partial z_{\bar\alpha_j}} &= \sum_{\alpha_1=0}^\infty\cdots\sum_{\alpha_m=0}^\infty (\alpha_j+1) p_{\alpha_1,\dots,\alpha_{j-1}-1,\alpha_j+1,\dots,\alpha_m} \prod_{k=1}^m z_{\bar\alpha_k}^{\alpha_k}
\end{align}
\end{widetext}
If we set $\bar\alpha_i=1$, limit $m=1$ and set the total number of new edges in the prefactor equal to $\alpha_j'=c$ then Eq \ref{eq:main-result} reduces to Eq 9 of Moore \textit{et al} \cite{Moore_Ghoshal_Newman_2006}.

\section{Uniform attachment}
\label{sec:uniform}

We now consider the specific case of uniform attachment across all degree classes (as opposed to preferential attachment kernels). Under uniform attachment, we have $\pi_{\bar\alpha_i,(\alpha_1, \dots, \alpha_m)} = 1$ for all $\bar\alpha_i$, meaning that new edges attach to vertices independently of their current degree. This simplification allows us to write a single generating function $g(z) = f_{\bar\alpha_i}(z)$ for all classes. Keeping $r$ general covers both growing networks ($r<1$), in which vertex addition outpaces deletion, and constant-size networks ($r=1$), in which the two are balanced; the constant-size case is recovered as a special limit at the end of the section. 

This section is structured as follows. In \ref{subsec:derivingg} we derive the generating function for arbitrary $r$ before extracting its coefficients to obtain the joint degree distribution in closed form in section \ref{subsec:coefficientsg}. We then find the marginal distributions in \ref{subsec:marginals} and the moments of the generating function in \ref{sec:moments}. We show that the tails of the marginals are strictly ordered due to the cascade of cliques under vertex removal in \ref{sec:genericity}. The clustering coefficient of the grown networks is derived in \ref{sec:clustering} and the giant component is found in \ref{sec:giant}. Finally, the percolation threshold is found in \ref{sec:percolation}, along with the finite-sized clusters in \ref{sec:finite-clusters} as well as the assortativity of the grown networks in \ref{sec:assortativity}.

\subsection{Deriving the generating function}
\label{subsec:derivingg}
Setting $f_{\bar\alpha_j}(z)=g(z)$ in the generating-function identity~\eqref{eq:main-result} and dividing through by $\bar\alpha_i$, the master equation reduces to a partial differential equation (PDE) of the form
\begin{equation} 
r\sum_{j=1}^{m} \gamma_j(z_{j-1} - z_j)\frac{\partial g}{\partial z_j} - \big(1 + K(z)\big)g(z) = - \prod_{j=1}^{m} z_j^{\alpha'_j},
\label{eq:pde-coupled}
\end{equation}
where we write $z_j\equiv z_{\bar\alpha_j}$ for brevity and adopt the convention that $z_0 \equiv 1$ (the degradation of an ordinary edge, $j=1$, simply removes it). The deletion rate $r$ multiplies the derivative (degradation) term, having entered directly from~\eqref{eq:main-result}; the constant-size case treated by setting $r=1$ is one point on this family. The kernel $K(z)$ is a linear function of the generating function variables:
\begin{equation}
    K(z) = \sum_{j=1}^{m} c_j (1 - z_j), \qquad \text{where} \quad c_j = \frac{\lambda_{i,j}}{\bar\alpha_i}.
    \label{eq:K-coupled}
\end{equation}
The crucial feature of Equation~\eqref{eq:pde-coupled} is the term $r\gamma_j(z_{j-1} - z_j)$ multiplying the partial derivative $\partial g / \partial z_j$. This coupling means that the evolution of the generating function with respect to $z_j$ depends on the value of $z_{j-1}$. Physically, this reflects the fact that when a clique of size $\bar\alpha_j$ loses a vertex, it degrades into a clique of size $\bar\alpha_{j-1}=\bar\alpha_j-1$. Thus, the dynamics of different clique classes are not independent but form a cascade of dependencies. To proceed we must convert the coupled differential system into a recursive linear chain of ODEs that can be solved.

To solve this first-order linear PDE, we employ the method of characteristics. We introduce a parameter $\sigma$ that parametrises curves in the $(z_1, \dots, z_m)$ space along which the PDE reduces to an ODE. The characteristic equations are obtained by matching the coefficients in the PDE. For $z_j$, we have
\begin{equation}
    \frac{dz_j}{d\sigma} = r\gamma_j(z_{j-1} - z_j), \qquad j = 1, \dots, m.
    \label{eq:char-z-coupled}
\end{equation}
For the generating function $g$ along the characteristics, the chain rule gives
\begin{align}
    \frac{dg}{d\sigma} =& \sum_{j=1}^{m} \frac{dz_j}{d\sigma} \frac{\partial g}{\partial z_j}\nonumber\\
    =& r\sum_{j=1}^{m} \gamma_j(z_{j-1} - z_j) \frac{\partial g}{\partial z_j}.
\end{align}
The factor $r$ multiplying the derivative sum is exactly that appearing in~\eqref{eq:pde-coupled}, so on substituting into~\eqref{eq:pde-coupled} it cancels and the equation for $g$ along the characteristics is independent of $r$,
\begin{equation}
    \frac{dg}{d\sigma} - \big(1 + K(z(\sigma))\big) g = -\prod_{j=1}^{m} z_j(\sigma)^{\alpha'_j}.
    \label{eq:char-g-coupled}
\end{equation}
We impose the initial conditions $z_j(0) = z_j$ (the point in $z$-space where we wish to evaluate $g$) and seek the solution as $\sigma \to \infty$.

The system~\eqref{eq:char-z-coupled} is a set of coupled linear ODEs. To solve it, we introduce the \emph{defect variables}
\begin{equation}
    u_j(\sigma) = 1 - z_j(\sigma), \qquad j = 1, \dots, m.
\end{equation}
Since $z_0 \equiv 1$, we have $u_0 \equiv 0$. Differentiating and substituting into Eq \eqref{eq:char-z-coupled}
\begin{align}
    \frac{du_j}{d\sigma} =& -\frac{dz_j}{d\sigma}\nonumber\\
    =& -r\gamma_j(z_{j-1} - z_j)\nonumber\\
    =& r\gamma_j(u_{j-1} - u_j).
\end{align}
Rearranging, we obtain the recurrence
\begin{equation}
    \frac{du_j}{d\sigma} + r\gamma_ju_j = r\gamma_ju_{j-1}, \qquad u_j(0) = 1 - z_j.
    \label{eq:u-recurrence}
\end{equation}
This recurrence describes a cascade where defects propagate from lower to higher clique topologies. The decay rates $\gamma_j$ depend on the number of edges in a $j$-clique and so larger cliques have a higher probability of being reduced. Intuitively this is because they have more edges and therefore random vertex removal will influence these cliques more than smaller ones. For the base case $j=1$, since $u_0 = 0$, the equation for $u_1$ is homogeneous
\begin{equation}
    \frac{du_1}{d\sigma} + r\gamma_1u_1 = 0, \qquad u_1(0) = 1 - z_1,
\end{equation}
and has the solution
\begin{equation}
    u_1(\sigma) = (1 - z_1) e^{-r\gamma_1\sigma}.
\end{equation}

For $j \geq 2$, Eq~\eqref{eq:u-recurrence} is an inhomogeneous first-order linear ODE with rate constant $r\gamma_j$. The general solution can be expressed as a linear superposition of the initial conditions propagated through the system. We write the solution in terms of the transition functions $\Psi_{j,k}(\sigma)$
\begin{equation}
    u_j(\sigma) = \sum_{k=1}^{j} (1 - z_k) \Psi_{j,k}(\sigma).
    \label{eq:u-general}
\end{equation}

Here, $\Psi_{j,k}(\sigma)$ represents the probability that a defect initially introduced at level $k$ has propagated to level $j$ at time $\sigma$ in a cascade process with rates $r\gamma_k, \dots, r\gamma_j$. The explicit form of $\Psi_{j,k}(\sigma)$ is found to be (see Appendix \ref{sec:appendix-transitions})
\begin{equation}
    {\Psi_{j,k}(\sigma) = \sum_{\ell=k}^{j} A_{j,k}^{(\ell)} e^{-r\gamma_\ell \sigma}}
    \label{eq:app-psi-main}
\end{equation}
where the coefficients $A_{j,k}^{(\ell)}$ are given by the recursion in the appendix. Notably these coefficients are independent of $r$: the deletion rate cancels in their defining relation (Appendix~\ref{sec:appendix-transitions}), so $r$ enters the transition functions only through the decay rates $r\gamma_\ell$, and the entire cascade solution for arbitrary $r$ is obtained from the constant-size case by the single replacement $\gamma_\ell\to r\gamma_\ell$. Transforming $u_j$ back to $z_j$, the characteristic curves are
\begin{equation}
    z_j(\sigma) = 1 - \sum_{k=1}^{j} (1 - z_k) \Psi_{j,k}(\sigma).
    \label{eq:z-sigma-sol}
\end{equation}
Note that as $\sigma \to \infty$, the defect variables $u_j(\sigma) \to 0$ due to the damping factors $e^{-r\gamma_\ell \sigma}$ inherent in $\Psi_{j,k}$, implying $z_j(\sigma) \to 1$.

We now compute $K(z(\sigma))$ by substituting~\eqref{eq:z-sigma-sol} into~\eqref{eq:K-coupled}
\begin{align}
    K(z(\sigma)) &= \sum_{j=1}^{m} c_j \big(1 - z_j(\sigma)\big) \nonumber\\
    &= \sum_{j=1}^{m} c_j \sum_{k=1}^{j} (1 - z_k) \Psi_{j,k}(\sigma).
\end{align}
Exchanging the order of summation (summing over $k$ first, then $j$ from $k$ to $m$)
\begin{align}
    K(z(\sigma)) &= \sum_{k=1}^{m} (1 - z_k) \sum_{j=k}^{m} c_j \Psi_{j,k}(\sigma) \nonumber\\
    &= \sum_{k=1}^{m} (1 - z_k) \mathcal{P}_k(\sigma),
    \label{eq:K-along-char}
\end{align}
where we define the generalized kernel polynomial
\begin{equation}
    \mathcal{P}_k(\sigma) = \sum_{j=k}^{m} c_j \Psi_{j,k}(\sigma).
    \label{eq:P-polynomial}
\end{equation}
Equation~\eqref{eq:char-g-coupled} is a first-order linear ODE of the form
\begin{equation}
    \frac{dg}{d\sigma} - \big(1 + K(z(\sigma))\big) g = -\prod_{j=1}^{m} z_j(\sigma)^{\alpha'_j}.
\end{equation}
This can be written in standard form $g' + P(\sigma) g = Q(\sigma)$ with $P(\sigma) = -1 - K(z(\sigma))$. The integrating factor is
\begin{align}
    \mu(\sigma) =& \exp\left(-\int_0^{\sigma} \big(1 + K(z(s))\big)\, ds\right)\nonumber\\
    =& e^{-\sigma} \exp\left(-\int_0^{\sigma} K(z(s))\, ds\right).
\end{align}
Define the cumulative defect function
\begin{equation}
    \Lambda(\sigma) = \int_0^{\sigma} K(z(s))\, ds.
\end{equation}
Using~\eqref{eq:K-along-char} this becomes
\begin{align}
    \Lambda(\sigma) &= \sum_{k=1}^{m} (1 - z_k) \mathcal{I}_k(\sigma),
    \label{eq:Lambda-def}
\end{align}
where the function $\mathcal{I}_k(\sigma)$ is the integral of the kernel component $\mathcal{P}_k(\sigma)$
\begin{align}
    \mathcal{I}_k(\sigma) &= \int_0^{\sigma} \mathcal{P}_k(s)\, ds \nonumber \\
    &= \sum_{j=k}^{m} c_j \int_0^{\sigma} \Psi_{j,k}(s)\, ds.
    \label{eq:I-def-explicit}
\end{align}
The functions $\mathcal{I}_k(\sigma)$ are independent of $z$ and determined entirely by the model parameters and the decay rates. Using these definitions we find the integrating factor to be $\mu(\sigma) = e^{-\sigma} e^{-\Lambda(\sigma)}$. Multiplying both sides of the ODE by $\mu(\sigma)$ and integrating from $\sigma = 0$ to $\sigma = \infty$ we find
\begin{equation}
    \Big[\mu(\sigma) g(\sigma)\Big]_0^{\infty} = -\int_0^{\infty} \mu(\sigma) \prod_{j=1}^{m} z_j(\sigma)^{\alpha'_j}\, d\sigma.
\end{equation}
As $\sigma \to \infty$: $z_j(\sigma) \to 1$, so the generating function approaches a finite value, but $\mu(\sigma) \to 0$ (due to the $e^{-\sigma}$ factor). Thus $\mu(\infty) g(\infty) = 0$.
At the other boundary $\sigma = 0$: $\mu(0) = 1$ and $g(z(0)) = g(z)$, the value we are after. Therefore we obtain
\begin{equation}
g(z) = \int_0^{\infty} e^{-\sigma} e^{-\Lambda(\sigma)} \prod_{j=1}^{m} z_j(\sigma)^{\alpha'_j}\, d\sigma,
\end{equation}
giving the final integral representation of $g(z)$ as
\begin{equation}
    g(z) = \int_0^{\infty} e^{-\sigma} e^{-\Lambda(\sigma)} \prod_{j=1}^{m} \left[ 1 - \sum_{k=1}^{j} (1 - z_k) \Psi_{j,k}(\sigma) \right]^{\alpha'_j} d\sigma.
    \label{eq:g-integral-final}
\end{equation}
Equation~\eqref{eq:g-integral-final} is the generating function for an arbitrary deletion rate $r$; the decay rates $r\gamma_\ell$ enter through the transition functions~\eqref{eq:app-psi-main} and the cumulative defect $\Lambda(\sigma)$.

\emph{Constant-size limit.} Setting $r=1$ leads to a constant-size network in which addition and deletion are exactly balanced. In this limit the transition functions reduce to $\Psi_{j,k}(\sigma)=\sum_{\ell=k}^{j}A_{j,k}^{(\ell)}e^{-\gamma_\ell\sigma}$ and~\eqref{eq:g-integral-final} is a balanced-turnover generating function. Restricting further to a single clique class of ordinary edges reproduces the single-vertex addition-deletion solution of Moore \textit{et al.}~\cite{Moore_Ghoshal_Newman_2006} (see Appendix~\ref{sec:appendix-mapping}). More generally, the growing regime $r<1$ is obtained for any $r$ below unity, so that a single solution spans the entire range of addition-dominated dynamics.

\subsection{Coefficients of $g$ via series expansion}
\label{subsec:coefficientsg}

To obtain the exact joint degree distribution 
$p_{\alpha_1,\ldots,\alpha_m}$, we must extract the coefficient of the
monomial $\prod_{k=1}^{m} z_k^{\alpha_k}$ from the generating function
integral in \eqref{eq:g-integral-final}. Let us begin by substituting \eqref{eq:Lambda-def} into the
exponential term $e^{-\Lambda(\sigma)}$ and separate the $z$-dependent
components:
\begin{align}
e^{-\Lambda(\sigma)}
&= \exp\!\left( - \sum_{k=1}^{m} (1 - z_k)\, \mathcal{I}_k(\sigma) \right)
\nonumber \\
&= \exp\!\left( - \sum_{k=1}^{m} \mathcal{I}_k(\sigma) 
+ \sum_{k=1}^{m} z_k\, \mathcal{I}_k(\sigma) \right)
\nonumber \\
&= e^{-\mathcal{I}_{\mathrm{tot}}(\sigma)} 
\prod_{k=1}^{m} \exp\!\big( z_k\, \mathcal{I}_k(\sigma) \big),
\end{align}
where $\mathcal{I}_{\mathrm{tot}}(\sigma) = \sum_{k=1}^{m}
\mathcal{I}_k(\sigma)$. Using the series expansion for the exponential
$e^{x} = \sum_{r=0}^{\infty} x^{r}/r!$, we obtain
\begin{equation}
e^{-\Lambda(\sigma)}
= e^{-\mathcal{I}_{\mathrm{tot}}(\sigma)}
\prod_{k=1}^{m}
\sum_{r_k=0}^{\infty}
\frac{\mathcal{I}_k(\sigma)^{r_k}}{r_k!}\, z_k^{r_k}.
\label{eq:exp-expansion-z-final}
\end{equation}
Physically, the index $r_k$ represents the number of class-$k$ cliques accumulated by a vertex through attachment events during network evolution.

We next focus on the product term in the integrand of
\eqref{eq:g-integral-final}. Let the factor inside the square brackets for
class $j$ be $\mathcal{F}_j(z,\sigma)$. Using the transition functions
$\Psi_{j,k}(\sigma)$ defined earlier, we regroup terms by powers of
$z_k$:
\begin{align}
\mathcal{F}_j(z,\sigma)
&= 1 - \sum_{k=1}^{j} \Psi_{j,k}(\sigma)\, (1 - z_k)
\nonumber \\
&= \left( 1 - \sum_{k=1}^{j} \Psi_{j,k}(\sigma) \right)
+ \sum_{k=1}^{j} \Psi_{j,k}(\sigma)\, z_k .
\end{align}
The first term is the probability that a clique that began in class $j$ has, by age $\sigma$, degraded out of the topology set entirely, its last remaining edge removed when the vertex's final partner within it was deleted.
We define this probability as
\begin{equation}
A_j(\sigma) = 1 - \sum_{k=1}^{j} \Psi_{j,k}(\sigma).
\end{equation}

Thus, the factor raised to the power $\alpha'_j$ becomes
\begin{equation}
\big[\mathcal{F}_j(z,\sigma)\big]^{\alpha'_j}
= \left[
A_j(\sigma) + \sum_{k=1}^{j} \Psi_{j,k}(\sigma) z_k
\right]^{\alpha'_j}.
\end{equation}
Applying the binomial theorem to the outer exponent and the multinomial
theorem to the inner sum gives
\begin{align}
\left[\mathcal{F}_j\right]^{\alpha'_j}
&= \sum_{n_j = 0}^{\alpha'_j}
\binom{\alpha'_j}{n_j}
A_j(\sigma)^{\alpha'_j - n_j}
\left( \sum_{k=1}^{j} \Psi_{j,k}(\sigma) z_k \right)^{n_j}.
\label{eq:prod-expansion-z-final}
\end{align}

The inner power expands as
\begin{equation}
\left( \sum_{k=1}^{j}
\Psi_{j,k}(\sigma)\, z_k \right)^{n_j}
=
\sum_{\substack{
\ell_1^{(j)} + \cdots + \ell_j^{(j)} = n_j \\
\ell_k^{(j)} \ge 0
}}
n_j!\,
\prod_{k=1}^{j}
\frac{ \left( \Psi_{j,k}(\sigma)\, z_k \right)^{\ell_k^{(j)}} }{\ell_k^{(j)}!}.
\label{eq:inner_sum}
\end{equation}
Here, the index $\ell_k^{(j)}$ denotes the number of cliques of original class $j$ that have degraded to class $k$ and survived until time $\sigma$.

Substituting \eqref{eq:inner_sum} into
\eqref{eq:prod-expansion-z-final}, and combining with
\eqref{eq:exp-expansion-z-final} inside the integral
\eqref{eq:g-integral-final}, we isolate the coefficient of
$\prod_{k=1}^{m} z_k^{\alpha_k}$. The exponents of $z_k$ arise from the
attachment indices $r_k$ and the degradation indices $\ell_k^{(j)}$. The
conservation constraint for the total number of class-$k$ cliques is
\begin{equation}
r_k + \sum_{j=k}^{m} \ell_k^{(j)}
= \alpha_k,
\qquad k = 1,\dots,m.
\label{eq:constraint-final}
\end{equation}
This fixes
\[
r_k = \alpha_k - \sum_{j=k}^{m} \ell_k^{(j)},
\]
which requires $\sum_{j=k}^{m} \ell_k^{(j)} \le \alpha_k$. The joint degree distribution is then
\begin{equation}
p_{\alpha_1,\ldots,\alpha_m}
=
\sum_{\{\ell_k^{(j)}\}}
\mathcal{C}(\ell)\,
\int_{0}^{\infty}
e^{-\sigma}
\, e^{-\mathcal{I}_{\mathrm{tot}}(\sigma)}\,
\Phi(\sigma,\ell)\, d\sigma,
\end{equation}
where the sum runs over all non-negative integers $\ell_k^{(j)}$
satisfying the constraints. The combinatorial weight is
\begin{equation}
\mathcal{C}(\ell)
=
\left( \prod_{k=1}^{m} \frac{1}{r_k!} \right)
\prod_{j=1}^{m}
\left[
\frac{\alpha'_j!}{(\alpha'_j - n_j)!}
\prod_{k=1}^{j} \frac{1}{\ell_k^{(j)}!}
\right],
\end{equation}
where $n_j = \sum_{k=1}^{j} \ell_k^{(j)}$. The time-dependent kernel is
\begin{equation}
\Phi(\sigma,\ell)
=
\left( \prod_{k=1}^{m}
\mathcal{I}_k(\sigma)^{r_k} \right)
\prod_{j=1}^{m}
\left[
A_j(\sigma)^{\alpha'_j - n_j}
\prod_{k=1}^{j}
\Psi_{j,k}(\sigma)^{\ell_k^{(j)}}
\right].
\end{equation}

\subsection{Marginals of the joint degree distribution}
\label{subsec:marginals}
\begin{figure}[t]
    \centering
    \includegraphics[width=0.495\textwidth]{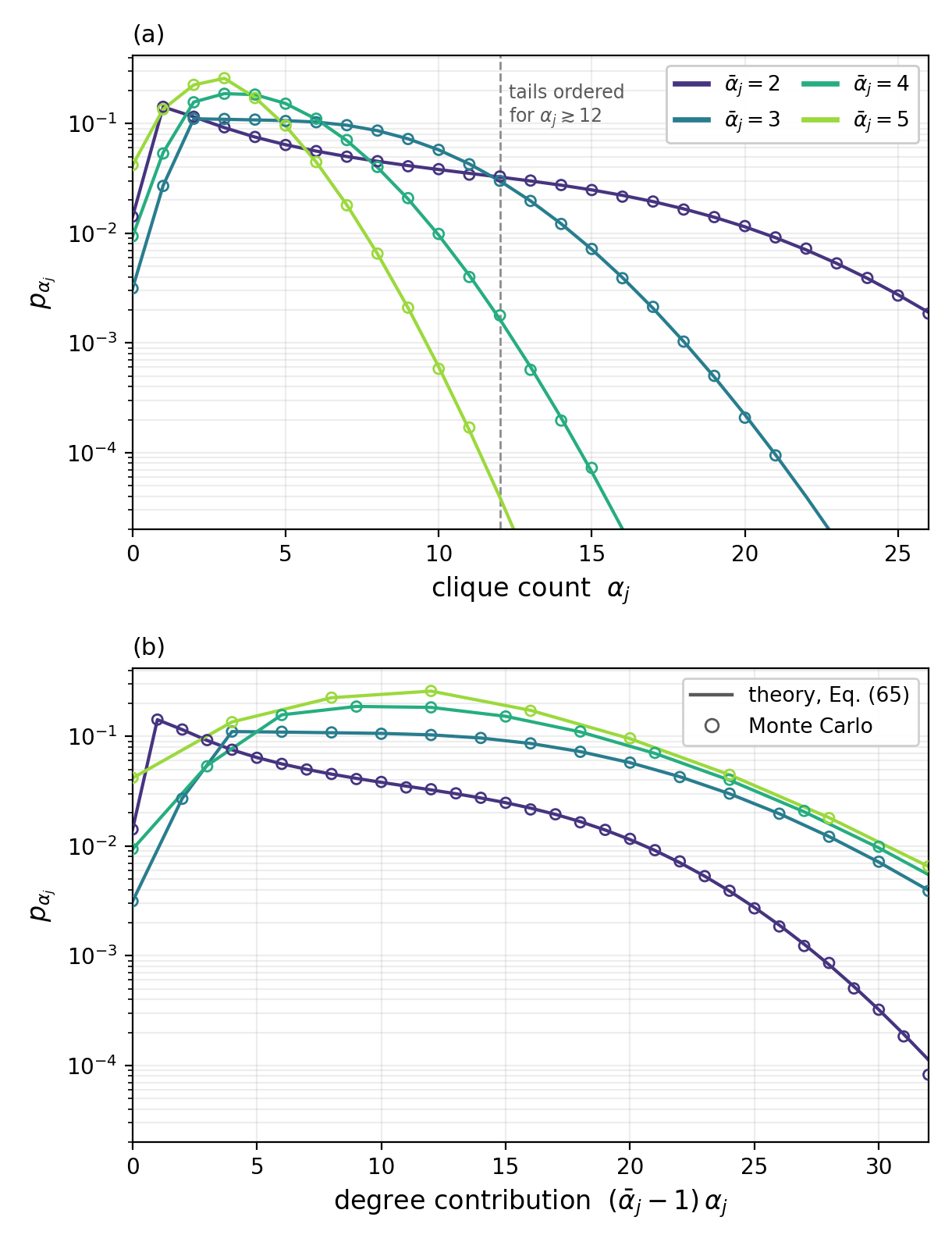}
    \caption{The marginal degree distributions for each clique size as a function of clique count, $\alpha_j$ (top). We also show the scaled results $(\bar\alpha_j-1)\alpha_j$, multiplying by the number of edges per-clique (bottom). Lines are the theoretical predictions from Eq \ref{eq:marginal-jdd} whilst scatter points are the average of Monte Carlo simulation. The parameters for the model are $\bm\chi=\{2,3,4,5\}$ and $(\alpha_1',\dots,\alpha_4')=(1,2,2,3)$ (one edge, two triangles, two 4-cliques and three 5-cliques per incoming vertex), with $\bar\alpha_i=4$. Simulations are the average of 20 independent realisations of networks with 12000 vertices, each evolved for 24000 addition--deletion steps.
    }    
    \label{fig:5-clique-sim}
\end{figure}

The marginal distribution $p_{\alpha_k}$ represents the probability that 
a randomly selected vertex belongs to exactly $\alpha_k$ cliques of size 
$\bar{\alpha}_k$, regardless of its membership in cliques of other sizes. 
This is extracted from the marginal generating function $g_k(z_k)$, defined as
\begin{equation}
    g_k(z_k) = g(1, \dots, z_k, \dots, 1),
\end{equation}
i.e. by setting all variables $z_j = 1$ for 
$j \neq k$ in the joint generating function $g(z)$. This operation sums the joint probability over all other clique dimensions.
Substituting this into the general integral representation 
\eqref{eq:g-integral-final}, all terms involving $(1 - z_j)$ vanish for 
$j \neq k$. Consequently, the cumulative defect kernel simplifies to
\begin{equation}
    \Lambda_k(\sigma) = (1 - z_k)\, \mathcal I_k(\sigma).
\end{equation}

Similarly, the product term in the integrand simplifies.  
For any clique topology $j < k$, the sum 
$\sum_{i=1}^{j} (1 - z_i)\Psi_{j,i}(\sigma)$ is zero because the index $i$ 
never equals $k$. For $j \ge k$, the only surviving term in the sum is the 
$i = k$ contribution. Therefore, the marginal generating function becomes
\begin{equation}
    g_k(z_k) = 
    \int_0^{\infty}
    e^{-\sigma}\,
    e^{-(1 - z_k)\mathcal I_k(\sigma)}
    \prod_{j=k}^{m}
        \left[
            1 - (1 - z_k)\Psi_{j,k}(\sigma)
        \right]^{\alpha'_j}
    d\sigma.
    \label{eq:g-marginal-integral}
\end{equation}
Physically, this implies that the membership count $\alpha_k$ (the number of class-$k$ cliques to which a vertex belongs) is populated by two 
mechanisms:  
(i) direct attachment of new vertices contributing cliques of class $k$ 
(the $\mathcal I_k$ process), and  
(ii) degradation of larger cliques of classes $j \ge k$ into class $k$ 
(the $\Psi_{j,k}$ transitions).  
Cliques initially smaller than class $k$ do not contribute to the marginal.

To extract the distribution $p_{\alpha_k}$, we expand the integrand in 
powers of $z_k$. The exponential term expands as
\begin{equation}
    e^{-(1 - z_k)\mathcal I_k(\sigma)}
    = e^{-\mathcal I_k(\sigma)}
      \sum_{r_k=0}^{\infty}
        \frac{\mathcal I_k(\sigma)^{r_k}}{r_k!}\,
        z_k^{r_k}.
\end{equation}
For each $j \ge k$, the product terms expand via the binomial theorem:
\begin{align}
    \left[
        1 - (1 - z_k)\Psi_{j,k}(\sigma)
    \right]^{\alpha'_j}
    =&
    \left[
        (1 - \Psi_{j,k}(\sigma))
        + z_k\,\Psi_{j,k}(\sigma)
    \right]^{\alpha'_j}
    \nonumber \\
    =&
    \sum_{l_j = 0}^{\alpha'_j}
        \binom{\alpha'_j}{l_j}
        (1 - \Psi_{j,k}(\sigma))^{\alpha'_j - l_j}\nonumber\\
        &\times\Psi_{j,k}(\sigma)^{l_j}
        z_k^{l_j}.
\end{align}
Here, $l_j$ represents the number of cliques of initial class $j$ that 
have degraded specifically into class $k$.

Combining these expansions, the coefficient of $z_k^{\alpha_k}$ is 
obtained by summing over all combinations of $r_k$ and 
$\{l_j\}_{j=k}^{m}$ satisfying the constraint
\[
    r_k + \sum_{j=k}^{m} l_j = \alpha_k,
\]
consistent with the notation of Eq.~\eqref{eq:constraint-final}; the attachment index $r_k$ is not to be confused with the deletion rate $r$.
The marginal degree distribution is therefore
\begin{align}
    p_{\alpha_k}
    =&
    \sum_{\substack{
        l_k,\dots,l_m \\
        0 \le l_j \le \alpha'_j
    }}
    \mathcal C_k
    \int_0^{\infty}
        e^{-\sigma}\,
        e^{-\mathcal I_k(\sigma)}
        \mathcal I_k(\sigma)^{r_k}\nonumber\\
        &\times\prod_{j=k}^{m}
            \left[
                (1 - \Psi_{j,k}(\sigma))^{\alpha'_j - l_j}
                \Psi_{j,k}(\sigma)^{l_j}
            \right]
    d\sigma,\label{eq:marginal-jdd}
\end{align}
where
\[
    r_k = \alpha_k - \sum_{j=k}^{m} l_j
    \qquad (r_k \ge 0),
\]
and the combinatorial factor is
\begin{equation}
    \mathcal C_k
    = 
    \frac{1}{r_k!}
    \prod_{j=k}^{m}
        \binom{\alpha'_j}{l_j}.
\end{equation}
We have plotted this expression in Fig \ref{fig:5-clique-sim}.

Equation~\eqref{eq:marginal-jdd} makes the departure from single-vertex dynamics explicit. At fixed age $\sigma$ the memberships received from younger cohorts are Poisson distributed with mean $\mathcal I_k(\sigma)$, whilst each of the $\alpha'_j$ cliques created at birth survives into class $k$ with probability $\Psi_{j,k}(\sigma)$, supplying the binomial factors; the stationary marginal is the $e^{-\sigma}$-weighted mixture of these over vertex age. The terms with $j>k$, memberships inherited from larger classes through degradation, have no counterpart in the single-vertex model of Ref.~\cite{Moore_Ghoshal_Newman_2006}, to whose constant-size, uniform-attachment degree distribution (expressible through incomplete gamma functions) the present solution collapses when the topology set is restricted to a single class of ordinary edges with single-vertex addition (Appendix~\ref{sec:appendix-mapping}). The cross-class binomial convolution in \eqref{eq:marginal-jdd} is therefore the analytic fingerprint of the topological cascade, and it is this structure that produces the inverted hierarchy of distribution widths discussed in Sec.~\ref{sec:discussion}.

\subsection{Moments of the joint degree distribution}
\label{sec:moments}

The clustering coefficient, the size of the giant component and the degree assortativity derived in the following sections are governed by the low-order moments of the joint degree distribution. These follow directly from the PDE~\eqref{eq:pde-coupled}, without requiring the full solution. Given their importance to subsequent sections, we will derive them here. Differentiating~\eqref{eq:pde-coupled} once with respect to $z_p$ and evaluating at $z=1$, with $g=1$ and $\partial g/\partial z_j=\langle\alpha_j\rangle$, gives the recurrence for the mean clique memberships,
\begin{equation}
(1+r\gamma_p)\langle\alpha_p\rangle - r\gamma_{p+1}\langle\alpha_{p+1}\rangle = \alpha'_p + c_p,
\label{eq:mean-recurrence}
\end{equation}
with the convention $\langle\alpha_{m+1}\rangle=0$. Solving from $p=m$ yields the closed form
\begin{equation}
\langle\alpha_p\rangle = \sum_{q=p}^{m}
\frac{r^{\,q-p}\prod_{i=p+1}^{q}\gamma_i}{\prod_{i=p}^{q}(1+r\gamma_i)}\,(\alpha'_q+c_q).
\label{eq:mean-closed}
\end{equation}
The second moments follow in the same manner. Writing $S_{pq}=\partial^2 g/\partial z_p\partial z_q\big|_{z=1}$ for the factorial second moments and differentiating~\eqref{eq:pde-coupled} twice at $z=1$,
\begin{equation}
\begin{split}
\big[1+r(\gamma_p+\gamma_q)\big]&S_{pq} - r\gamma_{q+1}S_{p,q+1} - r\gamma_{p+1}S_{p+1,q} \\
&= \alpha'_p\alpha'_q - \delta_{pq}\alpha'_p + c_q\langle\alpha_p\rangle + c_p\langle\alpha_q\rangle,
\end{split}
\label{eq:second-moment-recurrence}
\end{equation}
with $S_{p,m+1}=S_{m+1,q}=0$. This linear system is solved directly, after which the ordinary second moments are $\langle\alpha_p\alpha_q\rangle = S_{pq}+\delta_{pq}\langle\alpha_p\rangle$. Since the total degree is $k=\sum_j\gamma_j\alpha_j$, the first two degree moments are $\langle k\rangle=\sum_p\gamma_p\langle\alpha_p\rangle$ and $\langle k^2\rangle=\sum_{p,q}\gamma_p\gamma_q\langle\alpha_p\alpha_q\rangle$.

The degree assortativity of Sec.~\ref{sec:assortativity} calls for one order more, since its denominator contains the edge-end average $\langle k^2\rangle_e=\langle k^3\rangle/\langle k\rangle$ and hence the third degree moment. Writing $T_{pqu}=\partial^3 g/\partial z_p\partial z_q\partial z_u\big|_{z=1}$ for the factorial third moments and differentiating~\eqref{eq:pde-coupled} three times at $z=1$,
\begin{equation}
\begin{split}
\big[1+r(\gamma_p+\gamma_q&+\gamma_u)\big]T_{pqu}
- r\gamma_{p+1}T_{p+1,q,u} \\
&- r\gamma_{q+1}T_{p,q+1,u}
- r\gamma_{u+1}T_{p,q,u+1} \\
={}& \alpha'_p\alpha'_q\alpha'_u
- \delta_{pq}\,\alpha'_p\alpha'_u
- \delta_{qu}\,\alpha'_q\alpha'_p \\
&- \delta_{up}\,\alpha'_u\alpha'_q
+ 2\,\delta_{pq}\delta_{qu}\,\alpha'_p \\
&+ c_p S_{qu} + c_q S_{pu} + c_u S_{pq},
\end{split}
\label{eq:third-moment-recurrence}
\end{equation}
with $T_{pqu}=0$ whenever any index equals $m+1$. The structure repeats that of~\eqref{eq:second-moment-recurrence}: the degradation term of the PDE raises each index once, the source supplies the third factorial moment of the fixed injection (the Kronecker deltas converting powers of $\alpha'$ to falling factorials), and the kernel $K$ couples each attachment rate to the second moment of the remaining pair. The linear system is again solved directly, after which the ordinary third moments are
\begin{equation}
\begin{split}
\langle\alpha_p\alpha_q\alpha_u\rangle
= T_{pqu} &+ \delta_{pq}S_{pu} + \delta_{qu}S_{pq} + \delta_{up}S_{qp} \\
&+ \delta_{pq}\delta_{qu}\langle\alpha_p\rangle,
\end{split}
\label{eq:third-ordinary}
\end{equation}
and the third degree moment follows as
$\langle k^3\rangle=\sum_{p,q,u}\gamma_p\gamma_q\gamma_u\,\langle\alpha_p\alpha_q\alpha_u\rangle$.

Setting $r=1$ recovers the moments of the constant-size network, while under the single-vertex mapping of Appendix~\ref{sec:appendix-mapping} ($\bar\alpha_i=1$, $m=1$, $\alpha'_1=c$) Eq.~\eqref{eq:mean-closed} reduces to the mean degree $\langle k\rangle = 2c/(1+r)$ of Moore \textit{et al.}~\cite{Moore_Ghoshal_Newman_2006}.

\subsection{Universal ordering of the tails}
\label{sec:genericity}

The ordering of the marginal tails in Fig.~\ref{fig:5-clique-sim}, the
ordinary-edge distribution reaching furthest and the largest-clique distribution cutting off soonest, is not a feature of the particular injection vector used there. In this subsection we show that it holds for
\emph{every} admissible injection vector: although the precise value of each marginal and the shape of its low-$\alpha_k$ body depend on the fixed joint degree $(\alpha'_1,\dots,\alpha'_m)$ of the incoming cliques, the ordering of the tails is set by the degradation cascade, and hence by the topology set alone.

\begin{figure}[tb]
    \centering
    \includegraphics[width=\columnwidth]{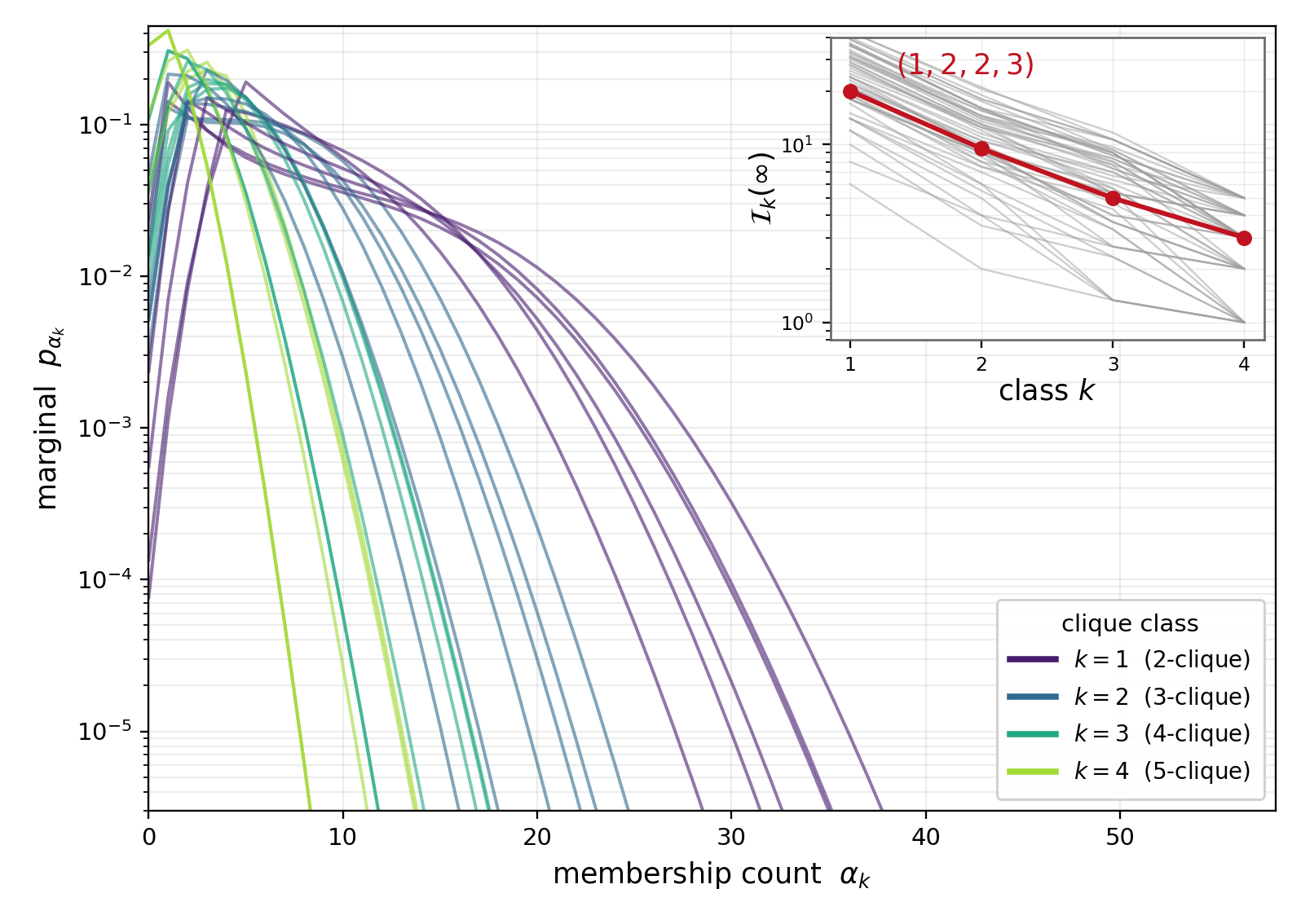}
    \caption{The tail ordering of the marginals is fixed by the topology set,
    not by the injection vector. \emph{Main panel:} marginal distributions
    $p_{\alpha_k}$ for the four clique classes of $\bm\chi=\{2,3,4,5\}$
    (colour), overlaid for six different injection vectors
    $(\alpha'_1,\dots,\alpha'_4)$ at $r=1$ and evaluated from the closed
    form~\eqref{eq:marginal-jdd}. The bodies scatter with the
    injection and each vector places the means differently, but at large
    $\alpha_k$ the distributions separate strictly by class, the ordinary edges ($k=1$) carrying the longest tail and the largest cliques ($k=4$)
    the sharpest cutoff, for every vector. \emph{Inset:} the tail cutoff
    $\mathcal{I}_k(\infty)=c_{\geq k}/(r\gamma_k)$ against class $k$ for many
    injection vectors drawn at random; every one is a strictly decreasing
    sequence [Eq.~\eqref{eq:Iinf-ordering}], with the vector of
    Fig.~\ref{fig:5-clique-sim} highlighted. The cutoffs themselves shift with
    the injection through $c_{\geq k}$, but their ordering does not.}
    \label{fig:tail-ordering}
\end{figure}

The transition functions $\Psi_{j,k}(\sigma)$ depend only on the topology set
and the deletion rate: the injection vector controls how many cliques of each
class are created, but not the fate of a clique once it exists. This
observation can be sharpened into an exact statement about the
marginals~\eqref{eq:marginal-jdd}. Because degradation proceeds one class at a
time, a clique created in class $j$ visits \emph{every} class $k\le j$ exactly
once on its way out of the topology set, residing in class $k$ for an
exponential time of rate $r\gamma_k$ (the clique is struck whenever one of its
$\gamma_k=\bar\alpha_k-1$ remaining partners is deleted). Since
$\Psi_{j,k}(\sigma)$ is the probability of occupying class $k$ at age
$\sigma$, its integral over all ages is the mean residence time,
\begin{equation}
    \int_0^\infty \Psi_{j,k}(\sigma)\,d\sigma \;=\; \frac{1}{r\gamma_k},
    \qquad k\le j,
    \label{eq:sojourn}
\end{equation}
independent of the class of birth; equivalently, from the closed
form~\eqref{eq:app-psi-main}, $\sum_{\ell=k}^{j}A^{(\ell)}_{j,k}/\gamma_\ell
= 1/\gamma_k$. Substituting~\eqref{eq:sojourn} into the definition of
$\mathcal{I}_k$ gives the mean number of received class-$k$ memberships held
by the oldest vertices,
\begin{equation}
    \mathcal{I}_k(\infty)
    \;=\; \frac{1}{r\gamma_k}\sum_{j=k}^{m} c_j
    \;\equiv\; \frac{c_{\geq k}}{r\gamma_k},
    \label{eq:Iinf}
\end{equation}
where $c_{\geq k}=\sum_{j\ge k}c_j$ is the total rate at which a vertex receives
cliques of class $k$ or above (Appendix~\ref{sec:appendix-mp}), all of which
must pass through class $k$ as they degrade. Both factors in Equation~\eqref{eq:Iinf} are monotone in the
class label: the flux $c_{\geq k}$ grows and the residence time $1/(r\gamma_k)$
lengthens as $k$ decreases, and we find
\begin{equation}
    \mathcal{I}_1(\infty) > \mathcal{I}_2(\infty) > \dots >
    \mathcal{I}_m(\infty)
    \label{eq:Iinf-ordering}
\end{equation}
for every injection vector, strictly wherever any class at or above $k$ is
created.

The far tail of each marginal is governed by $\mathcal{I}_k(\infty)$: the
birth cliques in~\eqref{eq:marginal-jdd} contribute at most
$\sum_{j\ge k}\alpha'_j$ memberships, so large values of $\alpha_k$ are reached only through the Poisson channel, which is largest for the oldest vertices, and up to subexponential factors
$p_{\alpha_k}\sim\mathcal{I}_k(\infty)^{\alpha_k}/\alpha_k!$ as
$\alpha_k\to\infty$. Consequently, evaluated at any sufficiently
large membership count, the marginals are strictly ordered,
$p_{\alpha_1}>p_{\alpha_2}>\dots>p_{\alpha_m}$. The injection vector sets the body of each distribution, its mean, through the linear
map~\eqref{eq:mean-closed}, and its shape at low counts. The injection vector also fixes the cutoffs $\mathcal{I}_k(\infty)$ through the fluxes $c_{\geq k}$; but the \emph{ordering} of the tails, with the longest tail residing with the ordinary edges and the sharpest cutoff with the largest cliques, is invariant, forced by the injection-independent residence times $1/(r\gamma_k)$ whose monotonicity in $k$ the fluxes only reinforce. Figure~\ref{fig:tail-ordering} shows this directly: across six different injection vectors the bodies scatter while the tails separate cleanly by class, and the inset confirms that $\mathcal{I}_k(\infty)$ is a strictly decreasing sequence for every one of a large random sample of vectors. The hierarchy of Fig.~\ref{fig:5-clique-sim} is therefore generic, holding for every admissible choice of the $\alpha'_j$.

\subsection{Clustering Coefficient}
\label{sec:clustering}

The local clustering coefficient $ C_v$ of a vertex $v$ is defined as the ratio of the number of triangles containing $v$ to the number of possible triangles that could exist given the degree of $v$. In a graph composed of edge-disjoint cliques, triangles can only exist within the cliques themselves; no triangles are formed by edges spanning different cliques.

The total degree $k$ of vertex $v$ is the sum of its neighbours across all clique memberships:
\begin{equation}
    k = \sum_{\bar\alpha_j\in\bm\chi} \alpha_j(\bar\alpha_j - 1).
\end{equation}
The overall degree distribution $p_k$ can be recovered from the joint degree distribution
\begin{equation}
    p_k=\sum_{\alpha_1=0}^\infty\cdots\sum_{\alpha_m=0}^\infty p_{\alpha_1,\dots,\alpha_m}\,\delta_{k,\,\sum_{j=1}^{m}\alpha_j(\bar\alpha_j-1)}
\end{equation}
The number of triangles $n_{j,\Delta}$ that vertex $v$ contributes due to its membership in a single clique of class $j$ (size $\bar\alpha_j$) is the number of pairs of neighbours within that clique:
\begin{equation}
    n_{j,\Delta} =  \frac{(\bar\alpha_j-1)(\bar\alpha_j-2)}{2}.
\end{equation}
The total number of triangles $v_{\Delta}$ associated with vertex $v$ is the sum over all its clique memberships:
\begin{equation}
    v_{\Delta} = \sum_{\bar\alpha_j\in\bm \chi}\alpha_j n_{j,\Delta}.
\end{equation}
The total number of possible triangles (pairs of neighbours) is given by $\binom{k}{2}$. Thus, the clustering coefficient is:
\begin{equation}
     C_v = \frac{\sum\limits_{\bar\alpha_j\in\bm \chi} \alpha_j (\bar\alpha_j-1)(\bar\alpha_j-2)}{k(k-1)}.
\end{equation}
For a randomly selected vertex the global clustering coefficient (the network transitivity) $\mathcal{C}$ is the ratio of the mean number of triangles per vertex to the mean number of connected triples per vertex,
\begin{equation}
     \mathcal{C} = \frac{\sum_{\alpha_1}\cdots \sum_{\alpha_m} v_\Delta\, p_{\alpha_1,\dots,\alpha_m}}{\sum_k\binom{k}{2}p_k}
     = \frac{\langle v_\Delta\rangle}{\big\langle \binom{k}{2}\big\rangle}.
\end{equation}
Substituting $v_\Delta=\tfrac{1}{2}\sum_j \alpha_j(\bar\alpha_j-1)(\bar\alpha_j-2)$ and $\binom{k}{2}=\tfrac{1}{2}k(k-1)$ we find
\begin{equation}
\mathcal{C}(r) = \frac{\displaystyle\sum_{j=1}^{m}(\bar\alpha_j-1)(\bar\alpha_j-2)\langle\alpha_j\rangle}{\langle k^2\rangle - \langle k\rangle},
\label{eq:clustering-r}
\end{equation}
with $\langle\alpha_j\rangle$ from~\eqref{eq:mean-closed} and $\langle k^2\rangle$ from the second moments~\eqref{eq:second-moment-recurrence}. Figure~\ref{fig:structural-r}(a) shows that $\mathcal{C}(r)$ increases with the deletion rate: as $r$ grows the triangle content of the numerator falls, but the degree variance $\langle k^2\rangle-\langle k\rangle$ falls faster because turnover preferentially thins the high-degree vertices that dominate $\langle k^2\rangle$, so the network becomes proportionally more clustered. Monte Carlo estimates confirm the prediction.

\subsection{Giant connected component}
\label{sec:giant}

We now find the size of the largest connected component of the stationary state of the grown network under percolation. Growth correlates the degrees of adjacent vertices: every clique attaches an incoming vertex to older ones, and the vertices present longest carry the most cliques, so the degrees at the two ends of an edge are positively correlated \cite{PhysRevLett.89.208701, PhysRevE.67.026126}. The connectivity of the grown network therefore cannot be read from the joint degree distribution alone and there are various formulations to approach this for block graphs \cite{PhysRevE.105.044314, Mann_Fang_Dobson_2025}. 

In this section we will introduce a new approach to solve this analytically using message passing \cite{10.1098/rspa.2022.0774}. We note that the correlations are generated by the vertex age $\sigma$ that Eq.~\eqref{eq:g-integral-final} already resolves. That expression writes $g$ as an average, with density $e^{-\sigma}$, of the age-conditional clique-degree generating function
\begin{equation}
F(z,\sigma)=e^{-\Lambda(\sigma)}\prod_{j=1}^{m}z_j(\sigma)^{\alpha'_j},
\label{eq:age-gf}
\end{equation}
whose product factor counts the cliques a vertex forms at birth and whose Poisson factor $e^{-\Lambda(\sigma)}$ counts those it later receives from younger vertices. Because the network is a proper block graph (every biconnected component is a clique and no two cliques share more than one vertex), it is locally tree-like at the level of cliques (the factor graph is treelike), and conditioning on age renders the memberships of neighbouring vertices independent. The giant component then follows from this age-resolved message passing~\cite{PhysRevLett.113.208702,Karrer_Newman_2010,PhysRevE.68.036112,Mann_Smith_Mitchell_Dobson_2021}.

\begin{figure*}[t]
    \centering
    \includegraphics[width=0.95\textwidth]{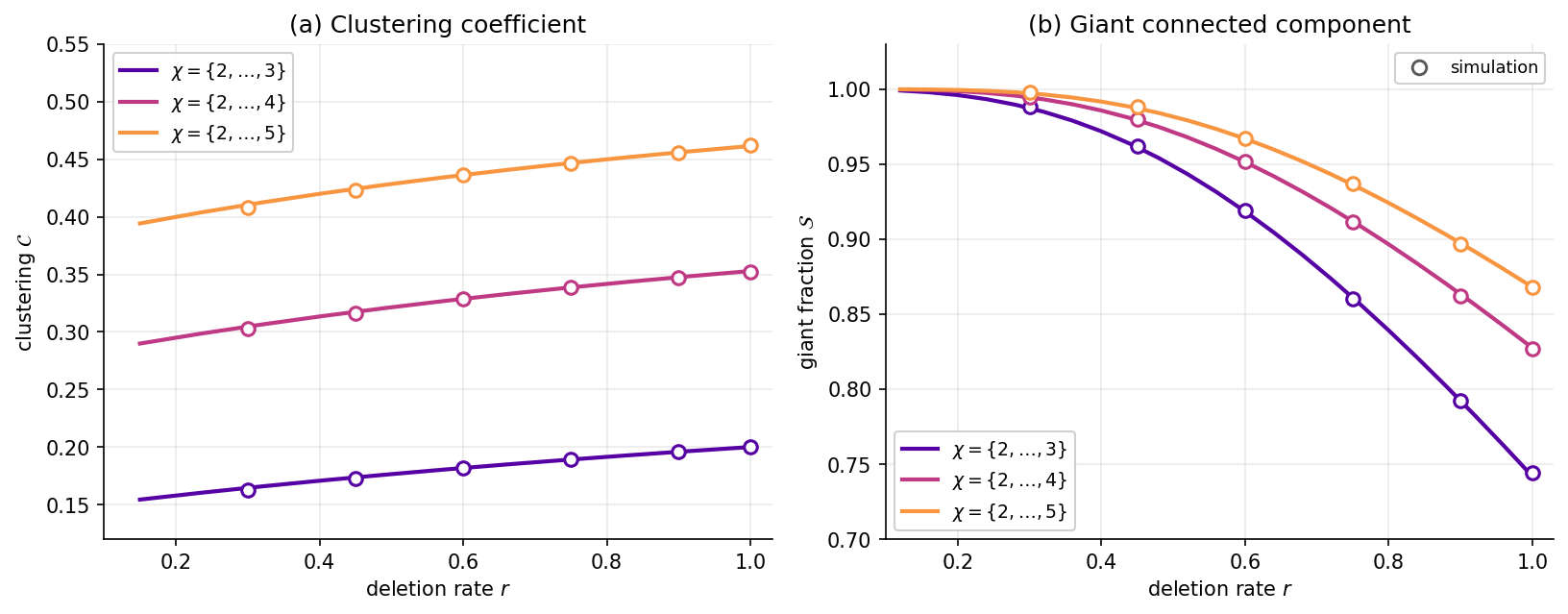}
    \caption{Structural properties as a function of the deletion rate $r$ for the consecutive topology sets $\bm\chi=\{2,\dots,n\}$ ($n=3,4,5$), in which each vertex joins one edge and one clique of the largest size $n$. Solid lines are the analytic predictions and circles are Monte Carlo estimates for the grown network, with error bars smaller than the markers. (a) Global clustering coefficient~\eqref{eq:clustering-r}, which rises both with $r$ and with clique size. (b) Giant component fraction from the age-resolved message passing~\eqref{eq:mp-fixed}--\eqref{eq:mp-S}: the prediction matches the grown network across the whole range, and $\mathcal{S}$ decreases with $r$ as turnover thins the network. Monte Carlo points are means over independent realisations of $8\times10^4$-vertex networks.}
    \label{fig:structural-r}
\end{figure*}

A vertex of age $\sigma$ meets its neighbours through three channels, distinguished by the ages of the co-members; the governing kernels are derived in Appendix~\ref{sec:appendix-mp}. It shares one \emph{birth} clique of $\bar\alpha_i$ vertices of common age $\sigma$, each surviving to the present with probability $e^{-r\sigma}$. For each size $\bar\alpha_j$ it \emph{created} $n_j=\alpha'_j-\delta_{\bar\alpha_j,\bar\alpha_i}$ attachment cliques at birth, joining it to $\bar\alpha_j-1$ older vertices whose ages exceed $\sigma$ by independent exponential increments. And it \emph{received} attachment cliques from younger vertices, forming for each size $\bar\alpha_j$ a Poisson process of rate $\gamma_j n_j$ over its history, each such clique contributing one younger root and $\gamma_j-1$ further older members. Let $\mathcal{V}(\sigma)$ be the probability that a vertex of age $\sigma$ is not joined to the giant component through any of its cliques. A neighbour reached along a clique extends the giant component unless it is disconnected through its \emph{remaining} cliques, an event of probability $\mathcal{V}$ with the shared clique removed. Collecting the three channels,
\begin{align}
\mathcal{V}(\sigma)=&B(\sigma)\prod_{j}A(\sigma)^{\gamma_j n_j}\nonumber\\
&\,
\times\exp\!\bigg[-\sum_{j}\gamma_j n_j\!\int_0^{\sigma}\!\Big(1-R_j(\tau)\,A(\tau)^{\gamma_j-1}\Big)d\tau\bigg],
\label{eq:mp-fixed}
\end{align}
where the birth clique contributes
\begin{equation}
B(\sigma)=\Big[1-e^{-r\sigma}\big(1-\mathcal{V}(\sigma)/B(\sigma)\big)\Big]^{\bar\alpha_i-1},
\label{eq:mp-birth}
\end{equation}
an older neighbour reached through a created clique contributes
\begin{equation}
A(\sigma)=1-e^{-r\sigma}\Big(1-\!\int_0^\infty\! e^{-w}\,\mathcal{V}(\sigma+w)\,dw\Big),
\label{eq:mp-att}
\end{equation}
and the younger root of a received size-$\bar\alpha_j$ clique contributes $R_j(\tau)=1-e^{-r\tau}\big(1-\mathcal{V}(\tau)/A(\tau)^{\gamma_j}\big)$; each excess factor is $\mathcal{V}$ with the shared clique removed. Equation~\eqref{eq:mp-fixed} is solved by iteration on the half-line, and the giant component fraction is
\begin{equation}
\mathcal{S}=1-\int_0^{\infty}e^{-\sigma}\,\mathcal{V}(\sigma)\,d\sigma.
\label{eq:mp-S}
\end{equation}
For a single class of ordinary edges the birth and created channels collapse and~\eqref{eq:mp-fixed}--\eqref{eq:mp-S} reduce to the standard excess-degree percolation condition for a growing network. Figure~\ref{fig:structural-r}(b) compares $\mathcal{S}(r)$ with direct simulation of the grown network: the message passing reproduces the measured giant component across the entire range of turnover. The giant component contracts monotonically as the deletion rate dilutes the network and networks with larger clique topologies are more robust under percolation.

\subsection{Percolation threshold and robustness}
\label{sec:percolation}

The giant component vanishes when the only solution of~\eqref{eq:mp-fixed} is the trivial one $\mathcal{V}\equiv1$, equivalently when the leading eigenvalue of the message passing linearised about $\mathcal{V}=1$ drops to unity. We locate the transition numerically, as the value of the control parameter at which $\mathcal{S}\to0$.

\begin{figure*}[t]
    \centering
    \includegraphics[width=\textwidth]{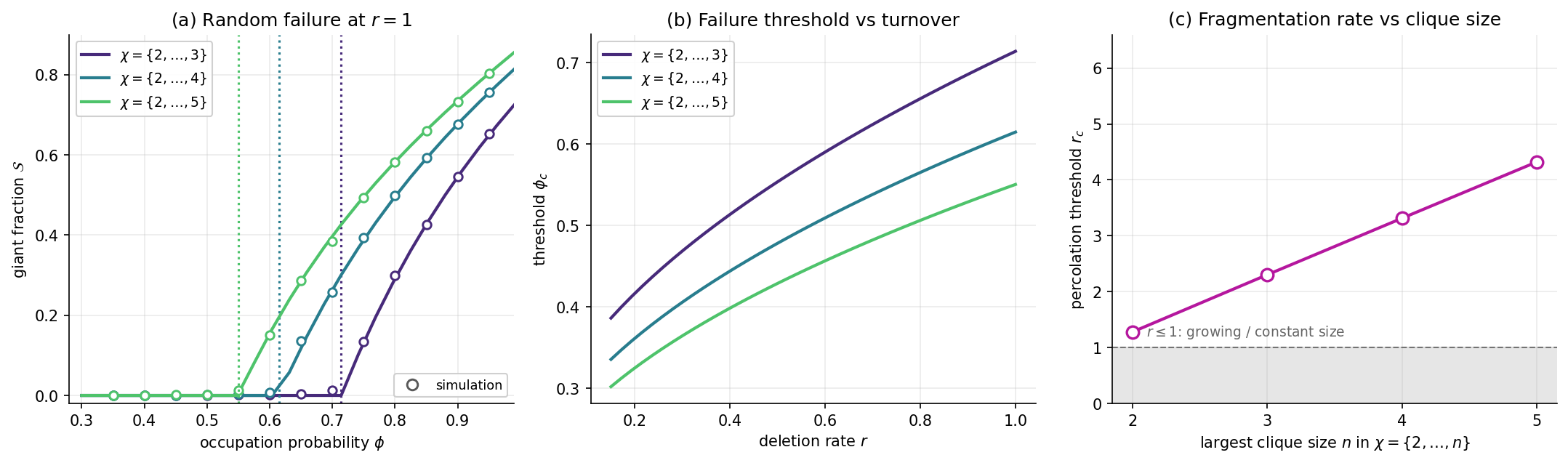}
    \caption{Percolation and robustness of the stationary network, for the consecutive topology sets $\bm\chi=\{2,\dots,n\}$ in which each vertex joins one edge and one clique of the largest size $n$. Solid curves are the age-resolved message passing and circles are simulation of the grown network. (a) Giant fraction $\mathcal{S}$ under random node failure as a function of the occupation probability $\phi$ at $r=1$; vertical dotted lines mark the message-passing thresholds $\phi_c$. (b) Random-failure threshold $\phi_c(r)$: the tolerance to damage falls (larger $\phi_c$) as the turnover rate $r$ increases, and is smallest for larger cliques. (c) Percolation threshold $r_c$ versus the largest clique size $n$ (the $n=2$ point is $\bm\chi=\{2\}$ with two edges per vertex). In every case $r_c>1$ and it rises with $n$: the shaded band $r\le1$, spanning all growing and constant-size networks, is entirely supercritical.}
    \label{fig:percolation}
\end{figure*}

Increasing the deletion rate thins the network, and for every consecutive topology set there is a threshold $r_c$ beyond which the stationary network carries no giant component [Fig.~\ref{fig:percolation}(c)]. In every case $r_c>1$, so turnover alone never fragments the network throughout the growing and constant-size regime $0<r\le1$. The threshold rises with clique size because a larger clique attaches to the rest of the network through more of its members, so more strongly clustered topologies are the more robust. The transition itself lies in the shrinking regime $r>1$, where the stationary solution is a formal continuation: direct simulation shows that a shrinking network collapses before its degree distribution relaxes to the stationary form, so $r_c>1$ marks the analytic boundary of the stationary solution rather than a transition realised by the dynamics. We leave it to future work to evaluate this regime analytically. 

The actual observed threshold is reached instead by external dilution. If each vertex of the stationary network is retained independently with probability $\phi$ under random node failure, a reached neighbour extends the giant component only when it is retained, which enters the message passing through the replacement $\mathcal{V}\to(1-\phi)+\phi\,\mathcal{V}$ in every excess factor; the fraction of the network occupied by the giant component then becomes
\begin{equation}
\mathcal{S}=\phi\Big(1-\int_0^{\infty}e^{-\sigma}\,\mathcal{V}(\sigma)\,d\sigma\Big),
\label{eq:perc-S}
\end{equation}
and a giant component survives only above an occupation threshold $\phi_c(r)$, again located where $\mathcal{S}\to0$. Figure~\ref{fig:percolation}(a) confirms the percolated solution against site-percolation simulation of the grown network across the whole transition. The threshold $\phi_c$ increases with the deletion rate [Fig.~\ref{fig:percolation}(b)]: although turnover cannot by itself disconnect the network, it steadily erodes the tolerance to damage, so a network sustaining faster turnover is fragmented by the removal of a smaller fraction of its vertices. Larger cliques lower $\phi_c$, so more strongly clustered networks are the most resilient to random failure.

\subsection{Finite clusters: the component-size distribution}
\label{sec:finite-clusters}

Below the percolation threshold every cluster is finite; above it a population
of finite clusters coexists with the giant component. Here we find the
distribution of finite-cluster sizes for an arbitrary injection vector
$(\alpha'_1,\dots,\alpha'_m)$ and examine the average component size above, at
and below the percolation threshold, deferring the derivation to
Appendix~\ref{app:finite-clusters}.

The calculation is organised around the observation, established in
Sec.~\ref{sec:giant}, that a vertex meets its neighbours through exactly three
kinds of clique, distinguished by when and how each clique formed relative to
the vertex's own birth. First, the \emph{birth clique}: the vertex arrives as
one of $\bar\alpha_i$ coeval vertices, and its
$\gamma_b=\bar\alpha_i-1$ co-members carry survival clocks that started with
its own. Second, the \emph{created cliques}: at birth the vertex founds
$n_j=\alpha'_j-\delta_{\bar\alpha_j,\bar\alpha_i}$ cliques of each size
$\bar\alpha_j$ (its prescribed memberships, less the birth clique when the
sizes coincide), each recruiting $\gamma_j=\bar\alpha_j-1$ \emph{older}
vertices chosen uniformly, so that
\begin{equation}
    \varphi=\sum_{j=1}^{m}\gamma_j n_j
    \label{eq:Gamma-def}
\end{equation}
counts the total number of older vertices met at birth. Third, the
\emph{received cliques}: for the rest of its life the vertex is itself
recruited by younger cohorts, per class a Poisson process of rate
$\gamma_j n_j$ in its own age. That the birth headcount $\varphi$ doubles as
the total reception rate is no accident: they are the same attachment slots,
viewed from the two ends of the resulting edges. Classes with $n_j=0$ drop
out of every formula below.

Let $\theta_s(\sigma)$ be the probability that a retained vertex of age $\sigma$
belongs to a finite component of size $s$, with generating function
\begin{equation}
    G(z,\sigma)=\sum_{s\ge1}\theta_s(\sigma)\,z^s ,
    \label{eq:gf-def}
\end{equation}
each power of $z$ counting one vertex of the component. The component
containing a vertex decomposes as the vertex itself together with the
sub-clusters reached through each of its cliques. Because any two cliques of
a block graph share only the vertex itself, and conditioning on age renders
the pieces independent (Sec.~\ref{sec:giant}), the sizes of the pieces add
and their generating functions multiply: $G$ carries one leading factor of
$z$ for the vertex and one multiplicative factor per clique channel. The
complication is that a neighbour's onward contribution depends on how it was
reached, since the shared clique must not be counted twice. The fixed point
therefore closes not on $G$ itself but on the \emph{cavity profiles}
$\mathbf v=(v_B,v_{C_1},\dots,v_{C_m},v_R)$, where $v_B$, $v_{C_j}$ and $v_R$
generate the cluster hanging off a vertex approached through, respectively,
its birth clique, a size-$\bar\alpha_j$ clique it created, and a clique it
received; that is, the corresponding channel is struck from its own
product. Appendix~\ref{app:finite-clusters} shows that
$G(z,\sigma)=v_R(z,\sigma)$ with
\begin{align}
    v_B &= z\,\mathcal A^{\varphi}\,\prod_{l=1}^{m}D^{(l)},
    \label{eq:vB}\\
    v_{C_j} &= z\,H_B\,\mathcal A^{\varphi-\gamma_j}\,
    \prod_{l=1}^{m}D^{(l)},
    \label{eq:vC}\\
    v_R &= z\,H_B\,\mathcal A^{\varphi}\,\prod_{l=1}^{m}D^{(l)} .
    \label{eq:vR}
\end{align}
Read against the full product $z\,H_B\,\mathcal
A^{\varphi}\prod_l D^{(l)}$, the three closures simply omit the struck
channel: $v_B$ lacks the birth factor $H_B$, $v_{C_j}$ lacks one created
factor $\mathcal A^{\gamma_j}$, and $v_R$ lacks nothing at all. The last is a
property of the Poisson process: removing a single point leaves its law
unchanged, so arriving through a received clique tells the vertex nothing
about its remaining received cliques. The same fact explains
$G=v_R$: a uniformly chosen vertex is statistically identical to one whose
received channel is short one clique.

The channel factors are each built from a single ``dead-or-alive'' bracket
per neighbour. The birth clique is closed by
$H_B=[1-\phi e^{-r\sigma}(1-v_B)]^{\gamma_b}$: a coeval co-member is
absent or unretained with probability $1-\phi e^{-r\sigma}$ (it must survive
the shared lifetime, $e^{-r\sigma}$, and pass the retention coin, $\phi$),
in which case it contributes a factor of one; otherwise it is present and
contributes its own onward cluster, generated by $v_B$ because the clique it
shares with the focal vertex is its birth clique. The $\gamma_b$ co-members
are independent, whence the power. The created and received channels are
both assembled from
\begin{align}
    \mathcal A(z,\sigma)&=1-\phi e^{-r\sigma}
    \big[1-(\mathsf J v_R)(\sigma)\big],
    \label{eq:calA}\\
    (\mathsf J g)(\sigma)&=\int_0^\infty e^{-w}\,g(\sigma+w)\,dw,
    \label{eq:Jop}
\end{align}
the single-older-neighbour factor, the $z$-resolved, site-percolated form of
$A(\sigma)$ of Sec.~\ref{sec:giant}. Its structure follows from the
enumeration of Appendix~\ref{sec:appendix-mp}: an older vertex recruited into a clique
was, at the moment of recruitment, a uniformly chosen vertex and hence of age
$w\sim\mathrm{Exp}(1)$; a time $\sigma$ later it is present with probability
$\phi e^{-r\sigma}$ and of age $\sigma+w$, and the operator $\mathsf J$
averages its onward reach over the unresolved increment $w$. Its onward
function is $v_R$ because, seen from the older vertex, the shared clique is
one it \emph{received}. A created clique of class $j$ joins the vertex to
$\gamma_j$ such older neighbours, one factor $\mathcal A$ apiece; the $n_j$
cliques of each class then supply the exponent $\varphi$ in
\eqref{eq:vB}--\eqref{eq:vR}. Finally, each received channel carries a
backward Volterra exponent,
\begin{equation}
    \partial_\sigma\ln D^{(j)}=-\gamma_j n_j\big[1-H^{(j)}_R(z,\sigma)\big],
    \qquad D^{(j)}(z,0)=1,
    \label{eq:volterra-ode}
\end{equation}
with single-clique factor
\begin{equation}
    H^{(j)}_R(z,\tau)=\big[1-\phi e^{-r\tau}\big(1-v_{C_j}(\tau)\big)\big]\,
    \mathcal A(z,\tau)^{\gamma_j-1}.
    \label{eq:Hj}
\end{equation}
Here $H^{(j)}_R$ is the contribution of one received clique whose root, the
younger vertex that recruited the focal vertex, is now aged $\tau$: the
leading bracket is the root's dead-or-alive bracket, with survival clock
$\tau$ running since the clique formed and onward function $v_{C_j}$ because
the shared clique is one the root \emph{created}, while the remaining
$\gamma_j-1$ members are older vertices recruited alongside the focal vertex
and supply one factor $\mathcal A(z,\tau)$ each. Since the received cliques
of class $j$ arrive as a Poisson process of rate $\gamma_j n_j$ over the
vertex's life, the whole collection contributes the exponential of the
integrated defect $1-H^{(j)}_R$; equation \eqref{eq:volterra-ode} is its
differential form, accumulating the vertex's recruitment history as it ages.

The cavities obey the algebraic links
\begin{equation}
    v_R=H_B\,v_B=\mathcal A^{\gamma_j}\,v_{C_j}\qquad(j=1,\dots,m),
    \label{eq:cavity-links}
\end{equation}
which state that striking a channel is division by its factor; a single
unknown profile therefore carries the fixed point once the exponents
\eqref{eq:volterra-ode} are adjoined. At $z=1$ every bracket collapses to the
substitution $V\to(1-\phi)+\phi V$ of Sec.~\ref{sec:giant}:
$v_R(1,\sigma)$ is the site-percolated $V(\sigma)$, so $G(1,\sigma)=1$ below
$\phi_c$, while above it
$\phi\int_0^\infty e^{-\sigma}[1-G(1,\sigma)]\,d\sigma=S$ recovers the giant
fraction, so the distribution is properly normalised on the finite clusters.

Equations \eqref{eq:vB}--\eqref{eq:vR} are a nonlinear two-point
boundary-value problem in $\sigma$, parametrised by $z$, and the two-point
character mirrors the process itself. The factor $\mathcal A(\sigma)$
requires $\mathsf J v_R$, the profile at all ages above $\sigma$ that
encodes the fates of a vertex's older contacts, and is integrated backward from
$\sigma=\infty$; the exponents \eqref{eq:volterra-ode} accumulate the
vertex's history of recruitment and run forward from $\sigma=0$; and for
$\gamma_j\ge2$ the forward integrand $H^{(j)}_R$ itself contains the backward
variable $\mathcal A$. This age coupling replaces the finite-dimensional
fixed point of the age-local configuration model, whose component
distribution Newman obtains in closed
form~\cite{Newman2007,PhysRevE.95.052303}, by one on a function space
(Appendix~\ref{app:finite-clusters}), and the distribution is instead built
constructively.

What survives is everything short of resummation. Below threshold the first
derivatives $\nu_a(\sigma)=\partial_z v_a(z,\sigma)|_{z=1}$ obey linear
equations. Differentiating \eqref{eq:vR} at $z=1$, where every channel
factor equals unity, gives
\begin{equation}
    \nu_R(\sigma)=1+\hat B(\sigma)+\sum_{j=1}^{m}n_j\,\hat C_j(\sigma)
    +\sum_{j=1}^{m}\hat R_j(\sigma),
    \label{eq:nuR-sum}
\end{equation}
which reads exactly as the process does: one for the vertex itself, plus the
expected number of vertices reached through the birth clique, through each
created clique, and through the received collection. The channel derivatives
are the brackets differentiated,
\begin{align}
    \hat B &= \gamma_b\,\phi e^{-r\sigma}\,\nu_B,
    \qquad
    \hat C_j = \gamma_j\,\phi e^{-r\sigma}\,(\mathsf J\nu_R),
    \label{eq:BC-def}\\
    \hat R_j &= \gamma_j n_j\,\phi\!\int_0^\sigma\! e^{-r\tau}
    \big[\nu_{C_j}(\tau)+(\gamma_j-1)(\mathsf J\nu_R)(\tau)\big]\,d\tau,
    \label{eq:R-def}
\end{align}
each a product of a headcount, a presence probability, and a mean onward
reach (the hats distinguishing these linearised channel means from the
factors $B(\sigma)$ and $R_j(\tau)$ of Sec.~\ref{sec:giant}): $\gamma_b$ co-members, alive with probability $\phi e^{-r\sigma}$,
each reaching $\nu_B$ further vertices; $\gamma_j$ older neighbours per
created clique, their reach age-averaged by $\mathsf J$; and, integrated over
the root ages $\tau$ of the received cliques, each live clique contributing
its root's reach $\nu_{C_j}$ together with the reaches of its $\gamma_j-1$
older members. The links \eqref{eq:cavity-links}, differentiated, close the
system through $\nu_B=\nu_R-\hat B$ and $\nu_{C_j}=\nu_R-\hat C_j$, leaving a single
linear forward--backward integral equation for $\nu_R$ whose age average is
the mean cluster size,
\begin{equation}
    \langle s\rangle=\int_0^\infty e^{-\sigma}\,\nu_R(\sigma)\,d\sigma,
    \label{eq:meansize}
\end{equation}
reproducing the susceptibility of Sec.~\ref{sec:giant}; above threshold the
derivative is taken about $v_R(1,\sigma)=V(\sigma)<1$.

The full distribution follows to any order. Writing
$v_a(z,\sigma)=\sum_{s\ge1}c^{(s)}_a(\sigma)\,z^s$ with
$\theta_s(\sigma)=c^{(s)}_R(\sigma)$, the fixed point delivers the coefficients
as the explicit recursion
\begin{equation}
    c^{(s)}_a(\sigma)=\big[z^{s-1}\big]\,
    \Psi_a\big(\mathbf v(z,\cdot)\,;\sigma\big),
    \label{eq:hierarchy}
\end{equation}
where $v_a=z\,\Psi_a(\mathbf v)$ abbreviates \eqref{eq:vB}--\eqref{eq:vR}.
The recursion is explicit because $z$ enters the closures only through the
single prefactor: a component of size $s$ is the vertex together with
sub-clusters whose sizes sum to $s-1$, each strictly smaller, so the
coefficient of $z^{s-1}$ in $\Psi_a$ involves only the profiles of orders
$p\le s-1$, already known on the whole half-line. Order $s$ is thus a
quadrature of lower-order profiles against fixed forward and backward
kernels (Appendix~\ref{app:finite-clusters}), no equation being inverted,
and $\theta_s=\int_0^\infty e^{-\sigma}\theta_s(\sigma)\,d\sigma$. The base order is closed,
\begin{align}
    \theta_1(\sigma)&=\big(1-\phi e^{-r\sigma}\big)^{\gamma_b+\varphi}
    \nonumber\\
    &\times\exp\Big(\!-\!\sum_{j=1}^{m}\gamma_j n_j\!\int_0^\sigma\!
    \big[1-(1-\phi e^{-r\tau})^{\gamma_j}\big]\,d\tau\Big),
    \label{eq:pi1}
\end{align}
and is read directly from the closures at $\mathbf v=0$: the vertex is
isolated exactly when all $\gamma_b+\varphi$ vertices met at birth are absent
or unretained, which supplies the prefactor, and every clique it has since received is
dead, each such clique requiring all $\gamma_j$ of its members (root and
older recruits alike) to be absent or unretained, whence the factor
$(1-\phi e^{-r\tau})^{\gamma_j}$ inside the exponent, which is elementary,
Eq.~\eqref{eq:pi1-integral}.

\begin{figure}
    \centering
    \includegraphics[width=0.5\textwidth]{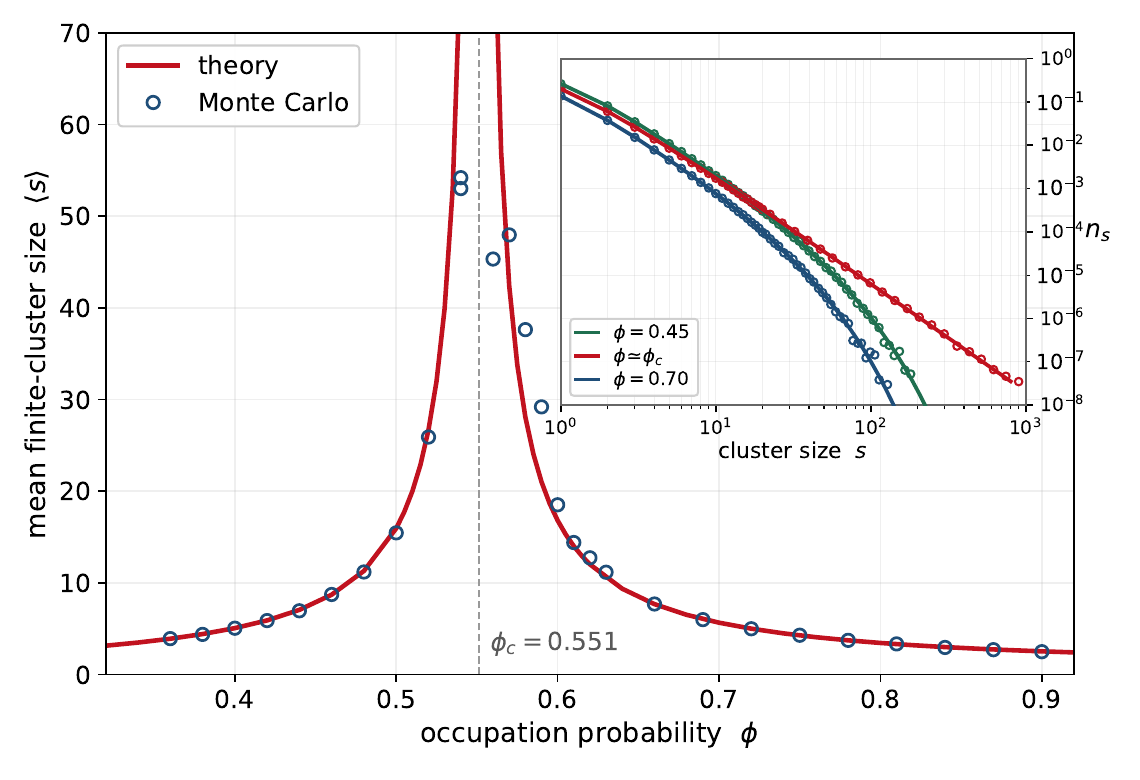}
    \caption{Finite clusters of the site-percolated network for
    $\bm\chi=\{2,3,4,5\}$ with one edge and one $5$-clique per incoming
    vertex, at $r=1$. Main: the mean size
    $\langle s\rangle$ of the finite clusters, Eq.~\eqref{eq:meansize},
    against the occupation probability $\phi$; the mean size diverges at
    the threshold $\phi_c\approx0.551$. Inset: the cluster-size density
    $n_s=\theta_s/s$ from the hierarchy \eqref{eq:hierarchy} below, at, and
    above the threshold (lines), against simulation. The decay is exponential on
    either side of $\phi_c$ and a power law at the threshold itself.}
    \label{fig:finite-clusters}
\end{figure}

The inset of Fig.~\ref{fig:finite-clusters} compares the resulting
cluster-size density $n_s=\theta_s/s$ with simulation across the transition for
the injection of one edge and one $5$-clique per incoming vertex. The
distribution decays exponentially on either side of $\phi_c$ and slows to a
power law at the threshold itself, where the branch point of the fixed point
in $z$ reaches $z=1$: this is the mechanism Newman identifies for the
configuration model~\cite{Newman2007}, carried through the age-resolved
fixed point, and it operates for every admissible injection, the
$(\alpha'_1,\dots,\alpha'_m)$ entering \eqref{eq:vB}--\eqref{eq:vR} only
through the exponents $\gamma_b$ and $\varphi-\gamma_j$ and the rates
$\gamma_j n_j$.

\subsection{Degree assortativity}
\label{sec:assortativity}

The message passing of Sec.~\ref{sec:giant} relied on vertex age as the hidden variable behind the degree correlations of a growing network: older vertices carry more cliques, and every edge ties together the ages of its two endpoints. In this section we use it to compute the degree assortativity of the network. We remark that the assortativity of random block graphs and edge-disjoint clique covered empirical networks have been studied previously \cite{Mann_Fang_Dobson_2025,PhysRevE.105.044314,PhysRevE.107.054303} using alternative generating function formulations.

The calculation rests on two age-resolved quantities, and both follow from the age-conditional generating function~\eqref{eq:age-gf} which generates the joint clique degree of a vertex of age $\sigma$. Differentiating ~\eqref{eq:age-gf} once at $z=\bm1$ gives the mean number of class-$j$ cliques that such a vertex belongs to,
\begin{equation}
    \langle\alpha_j\rangle_\sigma
    = \mathcal I_j(\sigma) + \sum_{l\ge j}\alpha'_l\,\Psi_{l,j}(\sigma).
    \label{eq:alpha-cond}
\end{equation}
The first term counts the cliques received from younger cohorts by age
$\sigma$. The second counts the birth cliques of class $l\ge j$ that have
survived, or degraded, into class $j$. Weighting each class by its degree
contribution $\gamma_j$ gives the mean degree at age $\sigma$,
\begin{equation}
    \bar k(\sigma) = \sum_{j}\gamma_j\,\langle\alpha_j\rangle_\sigma .
    \label{eq:kbar-age}
\end{equation}
Averaging this profile over the age density $e^{-\sigma}$ returns the overall
mean degree,
\begin{equation}
    \int_0^\infty e^{-\sigma}\,\bar k(\sigma)\,d\sigma = \langle k\rangle ,
    \label{eq:kbar-norm}
\end{equation}
which is a convenient check on~\eqref{eq:kbar-age}. The profile climbs
steadily with age. At birth it equals $\sum_j\gamma_j\alpha'_j$; for the oldest
vertices it saturates at $r^{-1}\sum_j\gamma_j^2 n_j$.

\begin{figure}[tb]
    \centering
    \includegraphics[width=\columnwidth]{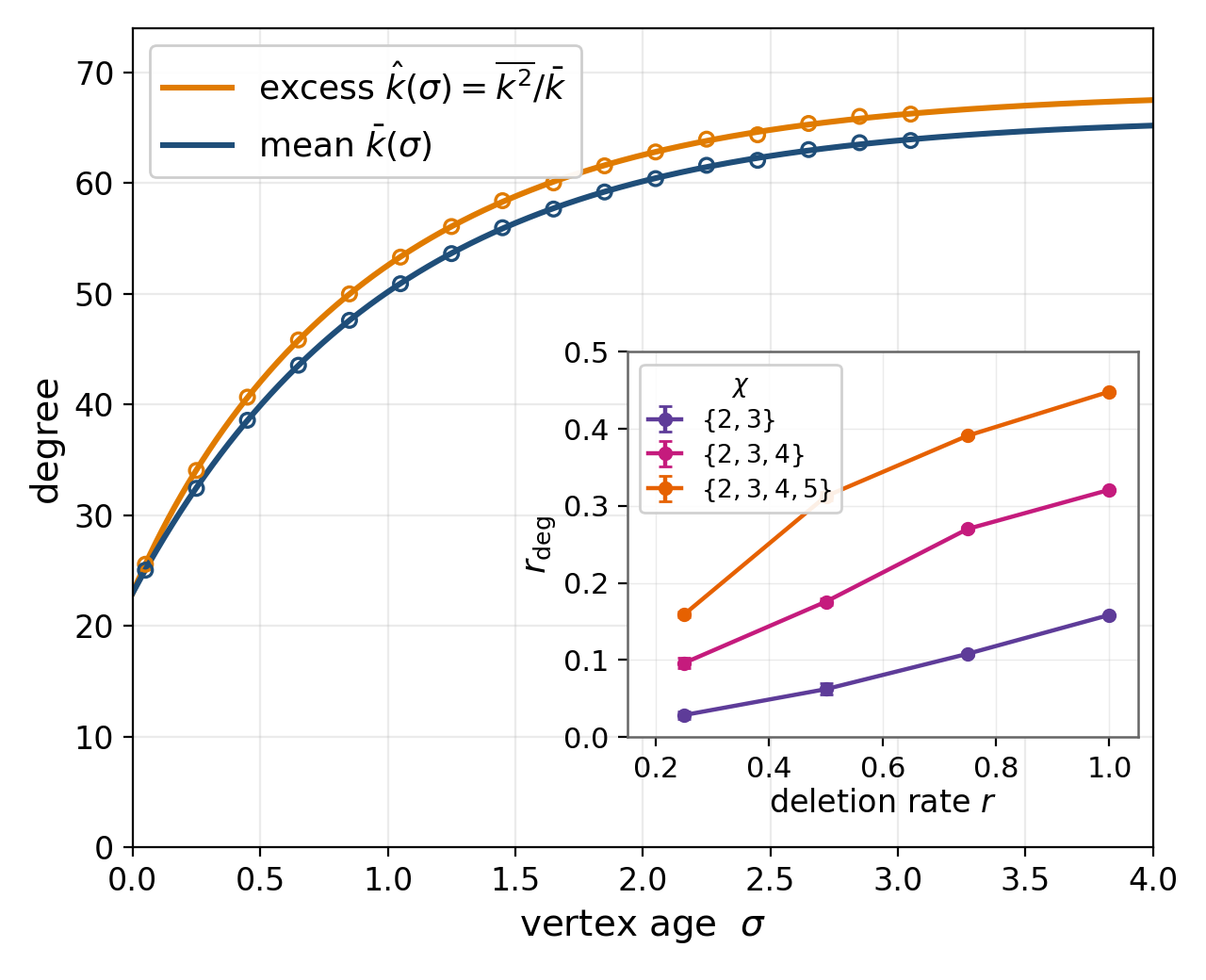}
    \caption{Degree assortativity of the grown network. \emph{Main panel:} the
    mean degree $\bar k(\sigma)$ [Eq.~\eqref{eq:kbar-age}] and the excess,
    edge-end degree $\hat k(\sigma)$ [Eq.~\eqref{eq:khat}] versus vertex age
    for $\bm\chi=\{2,3,4,5\}$ and $(\alpha'_1,\dots,\alpha'_4)=(1,2,2,3)$ at
    $r=1$ (lines), with Monte Carlo (circles). A vertex reached by traversing
    an edge is size-biased, so $\hat k(\sigma)>\bar k(\sigma)$. \emph{Inset:}
    Pearson degree assortativity $r_{\rm deg}$ versus deletion rate for the
    consecutive topology sets $\bm\chi=\{2,\dots,n\}$ with one edge and one
    clique of the largest size $n$ per incoming vertex. The network is
    positively assortative throughout, more strongly for larger cliques and
    higher turnover. Points are direct simulation of the grown network, with
    error bars giving the standard error over independent realisations.}
    \label{fig:assort}
\end{figure}

The mean degree is not, however, the degree one meets along an edge that has been selected at random. A vertex reached by following an edge tends to be one of high degree, simply because a high-degree vertex owns more edges through which it can be found. This is the familiar friendship paradox. The quantity we need is therefore the excess degree at age $\sigma$. It is the ratio of the
age-conditional second moment to the first,
\begin{equation}
    \hat k(\sigma) = \frac{\overline{k^2}(\sigma)}{\bar k(\sigma)} .
    \label{eq:khat}
\end{equation}
The two ingredients again come from $F$ in~\eqref{eq:age-gf}. Setting $z_k=e^{\gamma_k\theta}$ turns
$F$ into the moment generating function of the degree at fixed age; its second
cumulant is the age-conditional degree variance,
\begin{equation}
    \kappa_2(\sigma) = \sum_k\gamma_k^2\,\mathcal I_k(\sigma)
    + \sum_l\alpha'_l\Big[\textstyle\sum_{k\le l}\gamma_k^2\Psi_{l,k}
    - \big(\sum_{k\le l}\gamma_k\Psi_{l,k}\big)^{2}\Big].
    \label{eq:kappa2}
\end{equation}
The second moment is then $\overline{k^2}(\sigma)=\bar k(\sigma)^2+\kappa_2(\sigma)$,
so the excess degree exceeds the mean by the variance-to-mean ratio
$\kappa_2(\sigma)/\bar k(\sigma)$. Figure~\ref{fig:assort} plots both profiles
against simulation; the excess degree lies above the mean at every age.

With the two profiles in hand we turn to the assortativity itself. The Pearson correlation \cite{PhysRevLett.89.208701} of the degrees at the two ends of an edge is,
\begin{equation}
    r_{\rm deg}
    = \frac{\langle k\,k'\rangle_e-\langle k\rangle_e^{2}}
           {\langle k^2\rangle_e-\langle k\rangle_e^{2}} .
    \label{eq:assort-def}
\end{equation}
Here $\langle\cdot\rangle_e$ is an average over edge ends, and
$\langle k\,k'\rangle_e$ is the mean product of the two endpoint degrees. The
edge-end moments follow from the ordinary degree moments through
$\langle k^n\rangle_e=\langle k^{\,n+1}\rangle/\langle k\rangle$, and those
moments are available in closed form from the mean~\eqref{eq:mean-closed},
the second-moment recurrence~\eqref{eq:second-moment-recurrence}, and its
third-order extension~\eqref{eq:third-moment-recurrence}.

The network is positively assortative (inset, Fig.~\ref{fig:assort}). This
sets it apart from single-vertex growth models, in which the degrees at the
ends of an edge are usually anticorrelated \cite{Krapivsky_Redner_2001, PhysRevE.67.056104, PhysRevE.71.036127}. The effect grows with clique size:
at $r=1$ the coefficient rises from about $0.16$ for $\bm\chi=\{2,3\}$ to about
$0.45$ for $\bm\chi=\{2,3,4,5\}$. It also grows steadily with the deletion
rate. The mechanism is the clique cross-section that also drives the
clustering of Sec.~\ref{sec:clustering}. A larger clique binds the fates of its
members together, so the co-members of an old, high-degree vertex are
themselves old and of high degree. Turnover then sharpens the coupling,
because it strips away the youngest and lowest-degree attachments before they
can dilute it. Degree assortativity thus joins clustering and percolation
robustness as a structural fingerprint left by the cohort dynamics, one that
the age-resolved solution makes transparent through the
profiles~\eqref{eq:kbar-age} and~\eqref{eq:khat}.

\section{Discussion}
\label{sec:discussion}

In this paper, we have presented a framework for the time evolution of clique-clustered networks subject to simultaneous addition and deletion processes. While standard network growth models often focus on single-vertex addition, or treat clustering as a byproduct, our approach explicitly treats cliques (fully connected subgraphs) as the fundamental building blocks of the network topology. We focused specifically on block graphs, where cliques share vertices but are edge-disjoint, a structure that approximates many social and collaborative systems where groups form coherently.

The primary theoretical contribution of this work is the derivation and exact solution of the master equation for the joint degree distribution in the presence of higher-order topology degradation. We demonstrated that vertex deletion in a clique-based network introduces a non-trivial coupling between degree classes: the removal of a vertex from an $n$-clique does not merely remove edges, but converts the remaining topology into an $(n-1)$-clique. This creates a cascade of dependencies where the population of smaller cliques is continuously fed by the erosion of larger ones.

For uniform attachment at an arbitrary deletion rate $r$, we resolved this coupling by transforming the master equation into a partial differential equation (PDE). Along with the use of generating functions, we utilised the method of characteristics to map the coupled system onto a recursive chain of ordinary differential equations. This allowed us to derive the transition functions $\Psi_{j,k}(\sigma)$, which act as Green's functions describing the propagation through the hierarchy of clique sizes; the deletion rate enters these functions only through their decay rates $r\gamma_\ell$, so a single solution spans growing ($r<1$) and constant-size ($r=1$) networks alike. We recovered the constant-size, balanced-turnover network as the special case $r=1$, and the single-vertex solution of Moore \textit{et al.}~\cite{Moore_Ghoshal_Newman_2006} as a further limiting case restricted to 2-cliques. Furthermore, by extracting the exact coefficients of the generating function, we provided a closed-form expression for the asymptotic joint degree distribution and an analytical expression for the clustering coefficient. We then obtained the size of the giant connected component and the percolation threshold governing robustness to random vertex failure, exactly and across the full range of turnover. To achieve this we used message passing, where the vertex age acted as the hidden variable that generates the degree correlations of the grown network \cite{PhysRevE.68.036112}. 

Figure \ref{fig:5-clique-sim} (top) reveals a clear hierarchy in the breadth of the degree distributions, providing insight into the physical stability of the topological structures. Despite the incoming vertices carrying a higher initial load of 5-cliques ($\alpha'_4=3$) compared to ordinary edges ($\alpha'_1=1$), the stationary distribution for 5-cliques is the most tightly peaked, while the 2-clique distribution exhibits the longest tail.

This inversion is a direct consequence of the \textit{clique cross-section}. In our deletion process, the probability of a clique being struck (and subsequently degraded), scales linearly with its size, $\bar\alpha_j$. A 5-clique effectively presents a larger surface area to the deletion mechanism; it has five vulnerable nodes, any of which can trigger its collapse. Consequently, larger cliques are structurally fragile and have a short half-life within the network. They decay too rapidly for vertices to accumulate a large number of them, resulting in the sharp cutoff observed in their curve.

Conversely, the 2-cliques represent the most stable structures in the hierarchy. Not only do they possess the smallest cross-section for deletion, but they also act as the "ground state" of the topological cascade. As 5-cliques degrade to 4-, 3-, and finally 2-cliques, there is a constant probability flux flowing from higher to lower topologies. The 2-clique distribution therefore broadens due to two reinforcing effects: their inherent robustness against deletion and the accumulation of "debris" from the degradation of larger, more complex structures. This suggests that in networks subject to random node turnover, complex high-order motifs are transient, while simple dyadic interactions dominate the long-term structure.

This hierarchy is not an artefact of the particular injection vector $(\alpha'_1,\dots,\alpha'_4)=(1,2,2,3)$ used in Fig.~\ref{fig:5-clique-sim}; Sec.~\ref{sec:genericity} establishes that the ordering of the tails is fixed by the topology set alone. The closed form~\eqref{eq:mean-closed} makes the mechanism explicit: the largest class is suppressed by its own decay rate alone, $\langle\alpha_m\rangle=(\alpha'_m+c_m)/(1+r\gamma_m)$, whereas every class $p<m$ receives, in addition to its direct injection term $(\alpha'_p+c_p)/(1+r\gamma_p)$, a cascade contribution from every larger class [the $q>p$ terms of~\eqref{eq:mean-closed}]. Increasing the proportion of intermediate-size cliques therefore rescales the injection terms $\alpha'_q+c_q$, and with them the means, but cannot reorder the decay hierarchy; it instead increases the debris flux into every smaller class, broadening the low-order distributions further.

This structural hierarchy is further illuminated by rescaling the distributions by the edge-content of each clique, as shown in the bottom panel of Fig. \ref{fig:5-clique-sim}. Here, we plot the probability of a vertex having a specific \textit{partial degree} $k_j = \alpha_j(\bar{\alpha}_j-1)$ derived solely from topology $j$. The distributions for the higher-order cliques ($\bar\alpha \ge 3$) collapse almost on top of one another. This data collapse implies that while large cliques are statistically rarer due to their fragility, their high edge-density compensates exactly. A vertex is roughly equiprobable to sustain a degree contribution of $k=20$ through 5-cliques (requiring only 5 structures) as it is through 3-cliques (requiring 10 structures).

However, several open questions and avenues for future research remain. First, our stationary solution covers the growing and constant-size regimes ($r\le1$); the shrinking regime ($r>1$), where the percolation threshold $r_c$ formally lies (Sec.~\ref{sec:percolation}), remains open. A genuine treatment of net attrition likely requires following the finite-time, finite-size dynamics rather than a stationary distribution, and it is unclear whether the message-passing formulation introduced here can be adapted to that setting. 

Second, our exact solution relied on uniform attachment. In many real-world networks, attachment is preferential (e.g., linear or sub-linear preference). The attachment kernel could be generalised to allow different clique sizes to attach with different preferences, mimicking scenarios where larger groups are more selective in their connections. 

Third, we aim to relax the structural rigidity of the model. Currently, we assume the addition of pure cliques. Extending the formalism to arbitrary motifs (such as cycles, simplicial complexes, hypergraphs or neighbourhoods \cite{cantwell2019message}) would allow for the modelling of more diverse systems, though this would violate the block-graph assumption and require accounting for overlapping edges that our model currently excludes. It is hoped that this relaxation would provide an analytical framework for the generative growth of empirical networks \cite{PhysRevX.8.041011}. 

Finally, we believe that this work could help us derive a model of disease spreading with the vital dynamics of birth and death included. The challenge is coupling the addition-deletion model to the dynamics of the spreading process at each unit of time.

\appendix 

\section{Mapping to single vertices}
\label{sec:appendix-mapping}

In this appendix we show that the integral form of the generating function $g(z)$ in \eqref{eq:g-integral-final} reduces to the result found by Moore \textit{et al} in \cite{Moore_Ghoshal_Newman_2006} when we restrict the topology set to only include 2-cliques. To recover this, we set $\bm \chi =\{2\}$, $m=1$, $\bar\alpha_1=2$ and $r=1$ (the constant-size case). This implies that $\gamma_1=\bar\alpha_1-1=1$, and we set $\alpha_1'=c$ to map to the notation in \cite{Moore_Ghoshal_Newman_2006}. We have $\lambda(1,2,c)=c$ and hence $K(z)=c(1-z)$. The characteristic curve for ordinary edges $(j=1)$ is 
\begin{equation}
    z(\sigma) = 1-(1-z)e^{-\sigma},
\end{equation}
and the cumulative defect function $\Lambda(\sigma)$ becomes
\begin{align}
    \Lambda(\sigma) =& \int^\sigma_0c(1-z(s))ds\nonumber\\
    =&c(1-z)(1-e^{-\sigma}).
\end{align}
Gathering these results we write $g(z)$ as
\begin{equation}
    g(z)=\int^\infty_0 e^{-\sigma}e^{-c(1-z)(1-e^{-\sigma})}\Big[1-(1-z)e^{-\sigma}]^c d\sigma
\end{equation}
Substituting $u=(1-z)e^{-\sigma}$ we have
\begin{align}
    g(z)=&\int^0_{1-z}\Big(\frac{u}{1-z}\Big)e^{-c(1-z)+cu}(1-u)^c\Big(-\frac{du}{u}\Big),\nonumber\\
    =&\frac{e^{-c(1-z)}}{1-z}\int^{1-z}_0e^{cu}(1-u)^cdu
\end{align}
Setting $t=c(1-u)$ the integral can be written as
\begin{align}
    \int^{1-z}_0e^{cu}(1-u)^cdu=\frac{e^c}{c^{c+1}}\int^c_{cz}t^ce^{-t}dt
\end{align}
Next we use the upper incomplete Gamma function $\Gamma(a,x)=\int^\infty_xt^{a-1}e^{-t}dt$ to write 
\begin{align}
    \int^c_{cz}t^ce^{-t}dt=&\int^\infty_{cz}t^ce^{-t}dt- \int^\infty_{c}t^ce^{-t}dt\nonumber\\
    =& \Gamma(c+1,cz)-\Gamma(c+1,c)
\end{align}
Inserting this back into the expression for $g(z)$ we have 
\begin{equation}
    g(z)=\frac{e^{cz}}{1-z} c^{-(c+1)}\Big(\Gamma(c+1,cz)-\Gamma(c+1,c)\Big)
\end{equation}
which is Eq 11 from \cite{Moore_Ghoshal_Newman_2006}.

\section{Derivation of \texorpdfstring{$\Psi_{j,k}(\sigma)$}{Psi\_jk(sigma)}}
\label{sec:appendix-transitions}

Recall from Equation~\eqref{eq:u-recurrence} in the main text that the defect variables $u_j(\sigma)$ satisfy the recurrence (written here for $r=1$; the general deletion rate is restored at the end of this appendix)
\begin{equation}
    \frac{du_j}{d\sigma} + \gamma_j u_j = \gamma_j u_{j-1}, \qquad u_j(0) = 1 - z_j,
    \label{eq:app-u-recurrence}
\end{equation}
with boundary condition $u_0 \equiv 0$. We seek the transition functions $\Psi_{j,k}(\sigma)$ such that
\begin{equation}
    u_j(\sigma) = \sum_{k=1}^{j} (1 - z_k) \Psi_{j,k}(\sigma).
    \label{eq:app-u-expansion}
\end{equation}
The function $\Psi_{j,k}(\sigma)$ represents the contribution to the defect at level $j$ at time $\sigma$ arising from an initial unit defect at level $k$. It is the Green's function for the cascade from clique class $k$ to class $j$. By linearity, we can solve for each $\Psi_{j,k}$ independently by setting the initial condition $(1 - z_\ell) = \delta_{\ell k}$. Substituting the expansion~\eqref{eq:app-u-expansion} into the recurrence relation~\eqref{eq:app-u-recurrence}:
\begin{equation}
    \sum_{\ell=1}^{j}(1-z_\ell)\frac{d\Psi_{j,\ell}}{d\sigma} + \gamma_j \sum_{\ell=1}^{j}(1-z_\ell)\Psi_{j,\ell} = \gamma_j \sum_{\ell=1}^{j-1}(1-z_\ell)\Psi_{j-1,\ell}.
\end{equation}
Since this must hold for arbitrary initial conditions $(1-z_k)$, matching coefficients yields the system:
\begin{align}
    \text{For } k < j: \quad & \frac{d\Psi_{j,k}}{d\sigma} + \gamma_j \Psi_{j,k} = \gamma_j \Psi_{j-1,k}, \label{eq:app-psi-offdiag}\\[6pt]
    \text{For } k = j: \quad & \frac{d\Psi_{j,j}}{d\sigma} + \gamma_j \Psi_{j,j} = 0. \label{eq:app-psi-diag}
\end{align}
The initial conditions come from $u_j(0) = 1 - z_j$, which requires:
\begin{equation}
    \Psi_{j,k}(0) = \delta_{jk} = \begin{cases} 1 & \text{if } j = k, \\ 0 & \text{if } j \neq k. \end{cases}
    \label{eq:app-psi-ic}
\end{equation}

Since Eq.~\ref{eq:app-psi-diag} is diagonal, its solution is
\begin{equation}
    \Psi_{j,j}(\sigma) = e^{-\gamma_j \sigma}.
    \label{eq:app-psi-diag-sol}
\end{equation}
For $k < j$, we have an inhomogeneous equation driven by $\Psi_{j-1,k}$
\begin{equation}
    \frac{d\Psi_{j,k}}{d\sigma} + \gamma_j \Psi_{j,k} = \gamma_j \Psi_{j-1,k}, \qquad \Psi_{j,k}(0) = 0.
\end{equation}
Using the integrating factor $e^{\gamma_j \sigma}$
\begin{equation}
    \frac{d}{d\sigma}\left[e^{\gamma_j \sigma} \Psi_{j,k}\right] = \gamma_j e^{\gamma_j \sigma} \Psi_{j-1,k}.
\end{equation}
Integrating from $0$ to $\sigma$
\begin{equation}
    \Psi_{j,k}(\sigma) = \gamma_j e^{-\gamma_j \sigma} \int_0^{\sigma} e^{\gamma_j s} \Psi_{j-1,k}(s)\, ds.
    \label{eq:app-psi-recursion}
\end{equation}
This recursion, together with the diagonal solutions~\eqref{eq:app-psi-diag-sol}, fully determines all $\Psi_{j,k}$. We now derive a closed-form expression for the Green's functions. Starting from $\Psi_{k,k}(\sigma) = e^{-\gamma_k \sigma}$ and iterating the recursion, we claim:
\begin{equation}
    \Psi_{j,k}(\sigma) = \sum_{\ell=k}^{j} A_{j,k}^{(\ell)} e^{-\gamma_\ell \sigma},
    \label{eq:app-psi-ansatz}
\end{equation}
where the coefficients $A_{j,k}^{(\ell)}$ are to be determined.

We will verify this by substituting the ansatz into the recursion and examining the various constraints to arrive at a closed form expression for the coefficients. Substituting this ansatz into the recurrence~\eqref{eq:app-psi-offdiag} for $k < j$:
\begin{equation}
    \sum_{\ell=k}^{j} A_{j,k}^{(\ell)} (-\gamma_\ell) e^{-\gamma_\ell \sigma} + \gamma_j \sum_{\ell=k}^{j} A_{j,k}^{(\ell)} e^{-\gamma_\ell \sigma} = \gamma_j \sum_{\ell=k}^{j-1} A_{j-1,k}^{(\ell)} e^{-\gamma_\ell \sigma}.
\end{equation}
Matching coefficients of $e^{-\gamma_\ell \sigma}$: for $\ell = j$: $A_{j,k}^{(j)}(-\gamma_j + \gamma_j) = 0$, which is consistent for any $A_{j,k}^{(j)}$. For $k \leq \ell < j$
    \begin{equation}
        A_{j,k}^{(\ell)}(\gamma_j - \gamma_\ell) = \gamma_j A_{j-1,k}^{(\ell)}
        \label{eq:app-A-recursion}
    \end{equation}
which implies
\begin{equation}
   A_{j,k}^{(\ell)} = \frac{\gamma_j}{\gamma_j - \gamma_\ell} A_{j-1,k}^{(\ell)}.
\end{equation}
The initial condition $\Psi_{j,k}(0) = 0$ for $k < j$ requires
\begin{equation}
    \sum_{\ell=k}^{j} A_{j,k}^{(\ell)} = 0.
    \label{eq:app-A-constraint}
\end{equation}
Iterating the recursion~\eqref{eq:app-A-recursion} from level $\ell$ up to level $j$ expresses each coefficient through its diagonal value,
\begin{equation}
    A_{j,k}^{(\ell)} = \left(\prod_{i=\ell+1}^{j} \frac{\gamma_i}{\gamma_i - \gamma_\ell}\right) A_{\ell,k}^{(\ell)},
    \label{eq:app-A-iterate}
\end{equation}
and the diagonal values $A_{\ell,k}^{(\ell)}$ are themselves fixed by the initial condition $A_{k,k}^{(k)} = 1$ (from $\Psi_{k,k} = e^{-\gamma_k \sigma}$) together with the constraint~\eqref{eq:app-A-constraint} imposed at each level. This triangular system describes a linear cascade of first-order transitions, and its solution is the Bateman form
\begin{equation}
    A_{j,k}^{(\ell)} = \frac{\displaystyle\prod_{i=k+1}^{j}\gamma_i}{\displaystyle\prod_{\substack{i=k \\ i\neq \ell}}^{j}(\gamma_i-\gamma_\ell)}, \qquad k \leq \ell \leq j,
    \label{eq:app-A-explicit}
\end{equation}
a single expression that supplies every coefficient, the diagonal term $\ell=j$ included. One verifies directly that~\eqref{eq:app-A-explicit} satisfies both the recursion~\eqref{eq:app-A-recursion} and the constraint~\eqref{eq:app-A-constraint}, the latter because $\sum_{\ell=k}^{j}\prod_{i\neq\ell}(\gamma_i-\gamma_\ell)^{-1}=0$ for any set of $j-k+1\ge 2$ distinct rates. Substituting~\eqref{eq:app-A-explicit} into the ansatz gives the transition functions in closed form, matching Eq.~\ref{eq:app-psi-main} in the main text. We note that for the consecutive topology sets considered here the rates $\gamma_i=\bar\alpha_i-1$ are strictly increasing in the class label and hence pairwise distinct, so the denominators in~\eqref{eq:app-A-explicit} never vanish; degenerate (repeated) rates, which cannot arise for cliques of distinct sizes, would require the standard confluent limit of the Bateman solution.

For an arbitrary deletion rate $r$ the defect recurrence~\eqref{eq:app-u-recurrence} carries the rate $r\gamma_j$ in place of $\gamma_j$, so the diagonal solution becomes $\Psi_{j,j}(\sigma)=e^{-r\gamma_j\sigma}$ and the ansatz $\Psi_{j,k}(\sigma)=\sum_{\ell}A_{j,k}^{(\ell)}e^{-r\gamma_\ell\sigma}$. The coefficient recursion~\eqref{eq:app-A-recursion} then reads $A_{j,k}^{(\ell)}(r\gamma_j-r\gamma_\ell)=r\gamma_j A_{j-1,k}^{(\ell)}$, in which the common factor $r$ cancels. The coefficients $A_{j,k}^{(\ell)}$ are therefore exactly those of Eq.~\eqref{eq:app-A-explicit}, independent of $r$, and the deletion rate enters the transition functions solely through the exponential decay rates $r\gamma_\ell$.

\section{Age-resolved message passing}
\label{sec:appendix-mp}

This appendix derives the message passing of Sec.~\ref{sec:giant}: the age coordinate and survival law on which it rests, the enumeration of a vertex's cliques by the ages of their members, and the connectivity recursion that closes on the single function $\mathcal{V}(\sigma)$.

It is convenient to measure a vertex's age in the coordinate $\sigma$ of Sec.~\ref{sec:uniform} rather than in elapsed steps. Let $N$ be the current number of vertices and let $\sigma$ advance by $d\sigma=\bar\alpha_i/N$ per step, one step adding a cohort of $\bar\alpha_i$ vertices. A vertex is then deleted with probability $r\bar\alpha_i/N=r\,d\sigma$ per step, so it survives to age $\sigma$ with probability
\begin{equation}
(1-r\,d\sigma)^{\sigma/d\sigma}\longrightarrow e^{-r\sigma},
\label{eq:app-survival}
\end{equation}
the deletion acting at the single-vertex rate $r$. Since $\bar\alpha_i$ vertices are born per step, that is $N$ per unit $\sigma$, the population obeys $dN/d\sigma=(1-r)N$ and grows as $e^{(1-r)\sigma}$. The number of vertices of age in $[\sigma,\sigma+d\sigma]$ is the number born a coordinate-time $\sigma$ earlier, in proportion $e^{-(1-r)\sigma}$, weighted by their survival $e^{-r\sigma}$, so a uniformly chosen vertex has age density
\begin{equation}
\rho(\sigma)=e^{-\sigma},
\label{eq:app-agedensity}
\end{equation}
the weight already carried by Eq.~\eqref{eq:g-integral-final}. The construction below uses only \eqref{eq:app-survival} and \eqref{eq:app-agedensity}.

Fix a vertex of age $\sigma$. Its cliques fall into three families distinguished by the ages of their members. \emph{Birth clique.} The vertex belongs to exactly one clique of co-eval vertices, its birth cohort of size $\bar\alpha_i$; the $\bar\alpha_i-1$ co-members all have age $\sigma$ and each is present with probability $e^{-r\sigma}$. \emph{Created cliques.} At its birth the vertex forms $n_j=\alpha'_j-\delta_{\bar\alpha_j,\bar\alpha_i}$ cliques of each size $\bar\alpha_j$ by selecting $\bar\alpha_j-1$ existing vertices uniformly at random; the count $n_j$ is its $\alpha'_j$ memberships of that size, less the birth clique when $\bar\alpha_j=\bar\alpha_i$. Each co-member is older: chosen when the focal vertex had age $0$ it was a uniform vertex, of age $w\sim\mathrm{Exp}(1)$ by \eqref{eq:app-agedensity}, and a coordinate-time $\sigma$ later it is present with probability $e^{-r\sigma}$ and, if present, of age $\sigma+w$. \emph{Received cliques.} After its birth the vertex is itself selected as a target by younger vertices, at a rate fixed by counting slots: per unit $\sigma$ the $N$ incoming vertices each build $n_j$ size-$\bar\alpha_j$ cliques of $\bar\alpha_j-1$ targets, creating $N n_j(\bar\alpha_j-1)$ target slots, each filled by a uniform vertex and so assigned to a given vertex with probability $1/N$. The vertex thus acquires received size-$\bar\alpha_j$ cliques as a Poisson process of rate
\begin{equation}
\gamma_j n_j=n_j(\bar\alpha_j-1)
\label{eq:app-rate}
\end{equation}
in its own age, independent of $N$. This is precisely the per-vertex attachment rate $c_j=\lambda_{i,j}/\bar\alpha_i$ of Eq.~\eqref{eq:K-coupled}, of which the slot count is an independent check. A clique received when the vertex had age $\sigma-\tau$ was built by a vertex then of age $0$, so its root has current age $\tau$ and is present with probability $e^{-r\tau}$, while the root's other $\bar\alpha_j-2$ targets are older members of the form just described; as the rate \eqref{eq:app-rate} is constant, the elapsed ages $\tau$ of the received cliques are uniform on $[0,\sigma]$.

\begin{figure}
\centering
\begin{tikzpicture}[x=1cm,y=1cm,
  vert/.style={circle,draw,line width=0.7pt,minimum size=8pt,inner sep=0pt},
  focal/.style={vert,fill=black!15,draw=black!70,minimum size=10.5pt},
  bn/.style={vert,draw=violet!85!black,fill=violet!12},
  cn/.style={vert,draw=teal!65!black,fill=teal!12},
  rn/.style={vert,draw=orange!85!red,fill=orange!20},
  be/.style={violet!85!black,line width=0.7pt},
  ce/.style={teal!65!black,line width=0.7pt},
  re/.style={orange!85!red,line width=0.7pt},
  msg/.style={-{Stealth[length=5pt,width=4pt]},shorten >=1.5pt},
  lb/.style={font=\scriptsize,text=black!65,align=center,inner sep=1pt},
  mlb/.style={font=\scriptsize,inner sep=1pt},
  tt/.style={font=\scriptsize\bfseries,align=center,inner sep=1pt}]

\draw[-{Stealth[length=5pt]},black!60,line width=0.5pt] (0.30,0.75) -- (8.15,0.75);
\node[lb,anchor=north east] at (8.15,0.62) {vertex age};
\foreach \x/\l in {1.5/{$\tau$},4.2/{$\sigma$},6.2/{$\sigma{+}w$}}{
  \draw[black!60,line width=0.5pt] (\x,0.68)--(\x,0.82);
  \node[lb,anchor=north] at (\x,0.64) {\l};}
\draw[black!30,densely dashed,line width=0.4pt] (4.2,0.82)--(4.2,2.78);
\draw[black!30,densely dashed,line width=0.4pt] (6.2,0.82)--(6.2,3.72);

\node[focal] (F)  at (4.2,3.00) {};
\node[bn]    (bA) at (4.2,4.35) {};
\node[bn]    (bB) at (4.2,5.45) {};
\node[cn]    (mA) at (6.2,3.95) {};
\node[cn]    (mB) at (6.85,2.55) {};
\node[rn]    (rt) at (1.5,1.75) {};
\node[rn]    (ct) at (2.60,2.50) {};

\draw[be] (bA)--(bB);
\draw[be,msg] (bA)--(F);
\draw[be,msg] (bB) to[bend right=30] (F);
\draw[ce] (mA)--(mB);
\draw[ce,msg] (mA)--(F);
\draw[ce,msg] (mB)--(F);
\draw[re] (rt)--(ct);
\draw[re,msg] (rt)--(F);
\draw[re,msg] (ct)--(F);

\node[tt,violet!85!black] at (4.2,6.30) {birth clique};
\node[lb] at (4.2,6.00) {co-eval cohort, size $\bar\alpha_i$};
\node[lb,anchor=east,align=right] at (3.70,4.95) {co-members, age $\sigma$\\ present $e^{-r\sigma}$};
\node[mlb,violet!85!black,anchor=west] at (4.42,4.30) {$\mathcal V(\sigma)/B(\sigma)$};

\node[tt,teal!65!black,anchor=west] at (5.85,5.75) {created cliques};
\node[lb,anchor=west] at (5.85,5.45) {$n_j$ per class $j$};
\node[lb,anchor=west,align=left] at (6.42,4.42) {older members\\ age $\sigma{+}w$\\ present $e^{-r\sigma}$};
\node[mlb,teal!65!black,anchor=south] at (5.15,3.62) {$\mathcal V(\sigma{+}w)$};

\node[tt,orange!85!red,anchor=west] at (0.25,4.15) {received cliques};
\node[lb,anchor=west] at (0.25,3.85) {Poisson, rate $\gamma_j n_j$};
\node[lb,anchor=north,align=center] at (1.50,1.48) {root, age $\tau$\\ present $e^{-r\tau}$};
\node[mlb,orange!85!red,anchor=north west] at (2.78,2.10) {$\mathcal V(\tau)/A(\tau)^{\gamma_j}$};
\node[lb,anchor=south east,align=right] at (2.52,2.70) {co-target, age $\tau{+}w$\\ present $e^{-r\tau}$\\ \textcolor{orange!85!red}{passes $\mathcal V(\tau{+}w)$}};

\node[lb,anchor=north west,align=left,text=black!80] at (4.45,2.45) {focal vertex, age $\sigma$\\ computes $\mathcal V(\sigma)$};

\node[font=\footnotesize] at (4.30,0.02)
  {$\mathcal V(\sigma)=\textcolor{violet!85!black}{B(\sigma)}\,
    \textcolor{teal!65!black}{\textstyle\prod_j A(\sigma)^{\gamma_j n_j}}\,
    \textcolor{orange!85!red}{D(\sigma)}$};
\end{tikzpicture}
\caption{The three clique channels of a vertex of age $\sigma$, and the
message each neighbour passes. Members met at birth (the co-eval birth
cohort and the older vertices recruited into the created cliques) are
present with probability $e^{-r\sigma}$; a clique received at elapsed root
age $\tau$ carries its younger root and $\gamma_j-1$ older co-targets, each
present with probability $e^{-r\tau}$. A present neighbour passes the
probability that it fails to reach the giant component through its
\emph{remaining} cliques: its own $\mathcal V$, at its own age, with the
shared clique struck from the factorisation~\eqref{eq:app-factorise}
according to the role the clique plays \emph{for the neighbour}, which
reverses across the edge. A birth co-member strikes $B$; the root of a
received clique created it, and strikes one factor $A^{\gamma_j}$; a member
reached through a created clique received it, and strikes nothing, the
received cliques forming a Poisson process unchanged by the removal of one
point. The focal vertex converts each incoming message $m$ into the bracket
$1-e^{-r\,\cdot}(1-m)$ (absent, or present but failing) and multiplies the
brackets channel by channel into $B(\sigma)$, $A(\sigma)^{\gamma_j}$ and
$D(\sigma)$ of Eqs.~\eqref{eq:app-factorise}--\eqref{eq:app-D}.}
\label{fig:mp-channels}
\end{figure}
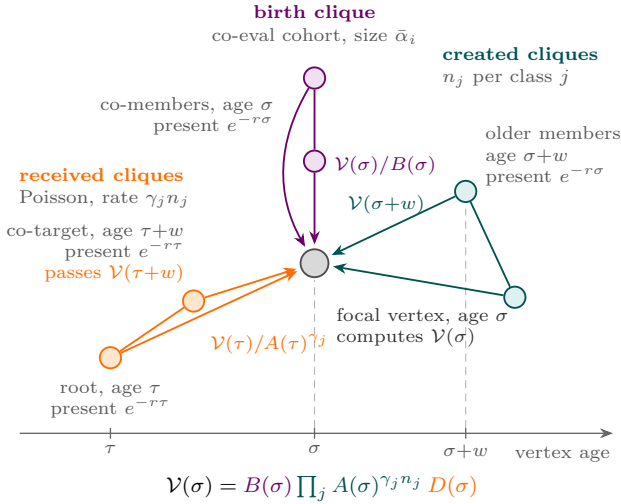

Because any two cliques share at most one vertex, the parts of the network reached through different cliques of a vertex meet only at that vertex and are otherwise disjoint; the age alone biases them, so conditioning on it makes them independent. Writing $\mathcal{V}(\sigma)$ for the probability that a vertex of age $\sigma$ reaches the giant component through none of its cliques, this independence factorises $\mathcal{V}$ over the three families,
\begin{equation}
\mathcal{V}(\sigma)=B(\sigma)\prod_j A(\sigma)^{\gamma_j n_j}D(\sigma),
\label{eq:app-factorise}
\end{equation}
where $B$, $A^{\gamma_j}$ and $D$ are the probabilities that the birth clique, one created size-$\bar\alpha_j$ clique, and the whole received collection fail to connect the vertex onward, Fig.~\ref{fig:mp-channels}. A clique connects the vertex to the giant when at least one of its other members does, and fails when every present member fails to connect through its \emph{own remaining} cliques; for a reached member this is an excess probability, $\mathcal{V}$ at that member's age with the shared clique struck from \eqref{eq:app-factorise}. Which factor is struck depends on the role the shared clique plays for the member, and that role is reversed across the edge: a clique the focal vertex created is one the older member received, and a clique it received is one its root created. The excess is thus $\mathcal{V}(\sigma)/B(\sigma)$ for a member reached through its birth clique, $\mathcal{V}(\sigma)/A(\sigma)^{\gamma_j}$ for one reached through a clique it created, and $\mathcal{V}(\sigma)$ for one reached through a clique it received. The last holds because the received cliques form a Poisson process, from which removing one clique leaves the remainder distributed as before.

The factors follow by collecting members. A single older neighbour, reached through a created or a received clique, is present with probability $e^{-r\sigma}$ and, arriving through a clique it received, carries the excess $\mathcal{V}(\sigma+w)$; averaging over its increment $w\sim\mathrm{Exp}(1)$ gives the single-neighbour factor $A(\sigma)$ of Eq.~\eqref{eq:mp-att}. The birth clique multiplies $\bar\alpha_i-1$ co-eval members, each present with probability $e^{-r\sigma}$ and reached through its own birth clique, giving $B(\sigma)$ of Eq.~\eqref{eq:mp-birth}; a created size-$\bar\alpha_j$ clique multiplies its $\bar\alpha_j-1$ older members, one factor $A(\sigma)$ each, failing with probability $A(\sigma)^{\gamma_j}$, and the vertex builds $n_j$ of them. A received size-$\bar\alpha_j$ clique with root age $\tau$ fails when its root and its $\bar\alpha_j-2$ older members all fail; the older members contribute $A(\tau)^{\gamma_j-1}$ and the root, present with probability $e^{-r\tau}$ and reached through a clique it created, contributes
\begin{equation}
R_j(\tau)=1-e^{-r\tau}\big(1-\mathcal{V}(\tau)/A(\tau)^{\gamma_j}\big),
\label{eq:app-Rj}
\end{equation}
so the clique fails with probability $R_j(\tau)A(\tau)^{\gamma_j-1}$. Summing over the received cliques, a Poisson process of rate $\gamma_j n_j$ in $\tau\in[0,\sigma]$, the received factor is
\begin{equation}
D(\sigma)=\exp\!\bigg[-\sum_j\gamma_j n_j\int_0^\sigma\big(1-R_j(\tau)A(\tau)^{\gamma_j-1}\big)\,d\tau\bigg],
\label{eq:app-D}
\end{equation}
the exponential being the probability that every clique of the process independently fails. Substituting $B$, $\prod_j A^{\gamma_j n_j}$ and \eqref{eq:app-D} into \eqref{eq:app-factorise} returns the self-consistent equation~\eqref{eq:mp-fixed}, and averaging $1-\mathcal{V}(\sigma)$ over \eqref{eq:app-agedensity} gives the giant fraction~\eqref{eq:mp-S}. Under random node failure a reached vertex extends the giant only when retained, which replaces each excess $\mathcal{V}$ by $(1-\phi)+\phi\mathcal{V}$ and yields Eq.~\eqref{eq:perc-S}.

The recursion of this appendix resolves a single event, whether a vertex
reaches the giant component, and $\mathcal V(\sigma)$ is accordingly a
number for each age. In the locally tree-like limit the complementary
statement is that the vertex belongs to a \emph{finite} component, so
$\mathcal V(\sigma)$ is the total probability mass of the finite-cluster
distribution, $\mathcal V(\sigma)=\sum_{s\ge1}\theta_s(\sigma)$. Appendix~%
\ref{app:finite-clusters} resolves that mass by cluster size, replacing each
scalar factor of the present appendix by a generating function in a variable
$z$ that counts vertices: the factors $B$, $A$ and $R_jA^{\gamma_j-1}$
constructed above reappear there as the $z=1$ values of $H_B$,
$\mathcal A$ and $H^{(j)}_R$, and the excess probabilities
$\mathcal V/B$, $\mathcal V/A^{\gamma_j}$ and $\mathcal V$ as the $z=1$
values of the cavity profiles $v_B$, $v_{C_j}$ and $v_R$. The reduction is
carried out explicitly in Appendix~\ref{app:z1-reduction}, and furnishes an
independent check of both constructions.

\section{Derivation of the finite-cluster distribution}
\label{app:finite-clusters}

This appendix derives the fixed point \eqref{eq:vB}--\eqref{eq:vR} of
Sec.~\ref{sec:finite-clusters} for an arbitrary injection vector
$(\alpha'_1,\dots,\alpha'_m)$, establishes the obstruction to a closed form
for $\theta_s$, and completes the moment equations and the coefficient
hierarchy quoted there. The argument follows the previous work of Newman \cite{Newman2007}, Kryven \cite{PhysRevE.95.052303} and Mann \& Dobson \cite{PhysRevE.107.054303} for the small components of clique clustered networks.

\subsection{The age-resolved generating function}

A vertex of age $\sigma$ is joined to the network through the three clique
families of Appendix~\ref{sec:appendix-mp}, with the member ages, presence
probabilities, received rates $\gamma_j n_j$, and conditional independence
of the reaches given age all established there; percolation additionally
retains every reached vertex independently with probability $\phi$.

Write $s_B$ for the number of vertices reached through the birth clique,
$s^{(j)}_a$ for the number reached through the $a$-th created clique of
class $j$ ($a=1,\dots,n_j$), and $t^{(j)}_\rho$ for the number reached
through the $\rho$-th received clique of class $j$, the received cliques of
class $j$ forming the Poisson set $\mathcal R_j$ with root ages
$\{\tau^{(j)}_\rho\}\subset[0,\sigma]$. A vertex sits in a component of size
$s$ exactly when these contributions sum to $s-1$. With one independent sum
for every clique the vertex belongs to,
\begin{widetext}
\begin{align}
    \theta_s(\sigma) =& \Bigg\langle\;
    \sum_{s_B\ge0}\;
    \prod_{j=1}^{m}\prod_{a=1}^{n_j}\sum_{s^{(j)}_a\ge0}\;
    \prod_{j=1}^{m}\prod_{\rho\in\mathcal R_j}\sum_{t^{(j)}_\rho\ge0}
    P_B(s_B)
    \Bigg[\prod_{j=1}^{m}\prod_{a=1}^{n_j}P^{(j)}_C\big(s^{(j)}_a\big)\Bigg]
    \Bigg[\prod_{j=1}^{m}\prod_{\rho\in\mathcal R_j}P^{(j)}_\rho\big(t^{(j)}_\rho\big)\Bigg]\nonumber\\
    &\times
    \delta\Bigg(s-1,\;s_B+\sum_{j=1}^{m}\sum_{a=1}^{n_j}s^{(j)}_a
    +\sum_{j=1}^{m}\sum_{\rho\in\mathcal R_j}t^{(j)}_\rho\Bigg)
    \Bigg\rangle_{\!\!\mathcal R},
    \label{eq:pi-delta}
\end{align}
the angle brackets denoting the average over the Poisson realisations
$\{\mathcal R_j,\{\tau^{(j)}_\rho\}\}$ of the received cliques and $\delta$
the Kronecker delta. This average runs over both the number of received
cliques and their root ages: for each class $j$ the set $\mathcal R_j$ holds
$N_j\sim\mathrm{Poisson}(\gamma_j n_j\,\sigma)$ cliques whose root ages are,
conditional on the count, independent and uniform on $[0,\sigma]$; this is the
order-statistics property of a homogeneous Poisson process, homogeneous here
because incoming vertices target the focal vertex at the constant rate
$\gamma_j n_j$ throughout its life. The classes are independent
processes. For any functional $F$ of the configurations, therefore,
\begin{equation}
    \big\langle F \big\rangle_{\!\mathcal R}
    \;=\;
    \prod_{j=1}^{m}\,\sum_{N_j=0}^{\infty}
    e^{-\gamma_j n_j\sigma}\,
    \frac{(\gamma_j n_j\sigma)^{N_j}}{N_j!}
    \int_0^{\sigma}\!\frac{d\tau^{(j)}_{1}}{\sigma}\cdots
    \int_0^{\sigma}\!\frac{d\tau^{(j)}_{N_j}}{\sigma}\;
    F\big(\{\tau^{(j)}_{\rho}\}\big),
    \label{eq:poisson-average}
\end{equation}
with the operator-product convention of \eqref{eq:pi-delta}. The remaining
randomness of a received clique (the presence and retention of its members,
their onward reaches, and the age increments $w$ of its older co-targets) is
private to that clique and is averaged pointwise, so each $\mathcal R_j$ is a
marked Poisson process whose marks are integrated out per point; this inner
average is what produces the single-clique factor $H^{(j)}_R$ of \eqref{eq:Hj},
and the outer average \eqref{eq:poisson-average} then acts on the product of
those factors. Multiplying \eqref{eq:pi-delta} by $z^s$ and summing over
$s\ge1$, the constraint lets the single power of $z$ distribute over the
cliques,
\begin{equation}
    z^{s} \;=\; z\;z^{s_B}
    \prod_{j=1}^{m}\prod_{a=1}^{n_j}z^{s^{(j)}_a}
    \prod_{j=1}^{m}\prod_{\rho\in\mathcal R_j}z^{t^{(j)}_\rho}
    \qquad\text{on the support of the delta},
    \label{eq:zs-factor}
\end{equation}
and the sum over $s$ then removes the delta, so that every convolution
collapses into a product of independent single-clique sums,
\begin{align}
    G(z,\sigma) &= z\,
    \Bigg[\sum_{s_B\ge0}P_B(s_B)\,z^{s_B}\Bigg]
    \prod_{j=1}^{m}\prod_{a=1}^{n_j}
    \Bigg[\sum_{s^{(j)}_a\ge0}P^{(j)}_C\big(s^{(j)}_a\big)\,z^{s^{(j)}_a}\Bigg]
    \Bigg\langle\prod_{j=1}^{m}\prod_{\rho\in\mathcal R_j}
    \sum_{t^{(j)}_\rho\ge0}P^{(j)}_\rho\big(t^{(j)}_\rho\big)\,z^{t^{(j)}_\rho}\Bigg\rangle_{\!\!\mathcal R}
    \label{eq:gf-factor}\\
    &= z\,H_B(z,\sigma)\,
    \prod_{j=1}^{m}\big[H^{(j)}_C(z,\sigma)\big]^{n_j}\,
    \prod_{j=1}^{m}D^{(j)}(z,\sigma),
    \label{eq:gf-product}
\end{align}
the created factors of a class being independent and identically distributed,
so that the $n_j$ single-clique sums of class $j$ supply the $n_j$-th power
of one channel generating function, while the independence of the classes in
\eqref{eq:poisson-average} splits the received average into
$\prod_j D^{(j)}$. Equation~\eqref{eq:gf-product} is the age-resolved
form of the product rule $G=z\prod_\tau H_{i\leftarrow\tau}$: the factors
$H_B$ and $H^{(j)}_C$ are the messages of single clique channels, while each
$D^{(j)}$ resums the single-clique messages $H^{(j)}_R$ of the Poisson
collection of received class-$j$ cliques; $G(z,\sigma)$ itself is defined in
Eq.~\eqref{eq:gf-def} of the main text.

\subsection{Closing the channels}

Each channel is now expanded through the vertices it reaches.

\paragraph*{Birth clique.}
$s_B$ is the sum over the $\gamma_b$ co-members of the vertices each of them
contributes. A co-member is absent or unretained with probability
$1-\phi e^{-r\sigma}$, and otherwise contributes itself and its onward reach
through its \emph{non-birth} cliques, with distribution $\tilde P_B$. Written
in full,
\begin{align}
    P_B(s_B) &= \sum_{s_1\ge0}\cdots\sum_{s_{\gamma_b}\ge0}
    \prod_{\mu=1}^{\gamma_b}
    \Big[(1-\phi e^{-r\sigma})\,\delta(s_\mu,0)
        + \phi e^{-r\sigma}\,\tilde P_B(s_\mu)\Big]\delta\Big(s_B,\;\textstyle\sum_{\mu=1}^{\gamma_b}s_\mu\Big).
    \label{eq:PB}
\end{align}
Multiplying by $z^{s_B}$ and summing, the delta again distributes the power
of $z$ over the members and the $\gamma_b$-fold sum factorises,
\begin{align}
    H_B(z,\sigma)
    &= \prod_{\mu=1}^{\gamma_b}\Big[(1-\phi e^{-r\sigma})
        + \phi e^{-r\sigma}\sum_{s_\mu\ge1}\tilde P_B(s_\mu)\,z^{s_\mu}\Big]
    \nonumber\\
    &= \big[\,1-\phi e^{-r\sigma}\big(1-z\,u_B(z,\sigma)\big)\big]^{\gamma_b}
    \nonumber\\
    &= \sum_{\ell=0}^{\gamma_b}\binom{\gamma_b}{\ell}
       \big(\phi e^{-r\sigma}\big)^{\ell}
       \big(1-\phi e^{-r\sigma}\big)^{\gamma_b-\ell}
       \big[z\,u_B(z,\sigma)\big]^{\ell},
    \label{eq:PhiB-full}
\end{align}
where $\sum_{s\ge1}\tilde P_B(s)\,z^{s}=z\,u_B(z,\sigma)$ defines the
co-member's onward (birth-excluded) generating function; the middle line is
the birth factor $H_B$ quoted after Eqs.~\eqref{eq:vB}--\eqref{eq:vR},
there written with $v_B=z\,u_B$, and the final binomial form displays the
sum over the number $\ell$ of co-members that are both present and retained.

\paragraph*{Created cliques.}
A created clique of class $j$ recruits $\gamma_j$ older members with
independent age increments; each is absent or unretained with probability
$1-\phi e^{-r\sigma}$, and otherwise contributes itself and its onward reach
at its own age $\sigma+w$,
\begin{align}
    P^{(j)}_C(s) &= \sum_{s_1\ge0}\cdots\sum_{s_{\gamma_j}\ge0}
    \prod_{\nu=1}^{\gamma_j}\Big[(1-\phi e^{-r\sigma})\,\delta(s_\nu,0)+ \phi e^{-r\sigma}\!\int_0^\infty\! e^{-w}\,\tilde P_R(s_\nu\!\mid\!\sigma+w)\,dw\Big]\,
    \delta\Big(s,\textstyle\sum_{\nu=1}^{\gamma_j}s_\nu\Big),
    \label{eq:PC}
\end{align}
with $\tilde P_R(\cdot\mid\sigma')$ the onward distribution of an older
member, which \emph{received} the shared clique. Generating as before and
writing $\sum_{s\ge1}\tilde P_R(s\mid\sigma')\,z^{s}=z\,u_R(z,\sigma')$,
every member supplies one factor
$1-\phi e^{-r\sigma}\big(1-z\int_0^\infty e^{-w}u_R(z,\sigma+w)\,dw\big)$,
which, since
$z\int_0^\infty e^{-w}u_R(z,\sigma+w)\,dw=(\mathsf J v_R)(\sigma)$ with
$v_R=z\,u_R$ and $\mathsf J$ of Eq.~\eqref{eq:Jop}, is precisely the
single-older-neighbour function $\mathcal A(z,\sigma)$ of
Eq.~\eqref{eq:calA}. Hence
\begin{equation}
    H^{(j)}_C(z,\sigma) = \mathcal A(z,\sigma)^{\gamma_j}.
    \label{eq:PhiC}
\end{equation}

\paragraph*{Received cliques.}
The received cliques of class $j$ form a Poisson process of rate
$\gamma_j n_j$ on $\tau\in[0,\sigma]$, and the average of a product over its
points resums exactly. Inserting the class-$j$ factor of
\eqref{eq:poisson-average} with $F=\prod_\rho H^{(j)}_R(z,\tau^{(j)}_\rho)$, the
$N$-fold integral factorises into identical single-age integrals,
\begin{align}
    D^{(j)}(z,\sigma)
    &= \Big\langle\prod_{\rho\in\mathcal R_j}H^{(j)}_R\big(z,\tau^{(j)}_\rho\big)\Big\rangle_{\!\mathcal R_j}
    \nonumber\\
    &= \sum_{N=0}^{\infty}e^{-\gamma_j n_j\sigma}\,
    \frac{(\gamma_j n_j\sigma)^{N}}{N!}\,
    \prod_{\rho=1}^{N}\int_0^\sigma\!\frac{d\tau_\rho}{\sigma}\,H^{(j)}_R(z,\tau_\rho)
    \nonumber\\
    &= \exp\Big(-\gamma_j n_j\!\int_0^\sigma\!
    \big[1-H^{(j)}_R(z,\tau)\big]\,d\tau\Big),
    \label{eq:PhiR}
\end{align}
with $H^{(j)}_R(z,\tau)$ the single-clique factor of Eq.~\eqref{eq:Hj}: a received
clique of root age $\tau$ contributes its root and its $\gamma_j-1$ remaining
older members. The root, which \emph{created} the clique, is present with
probability $e^{-r\tau}$ and carries the onward generating function
$u^{(j)}_C$ that excludes one created clique of class $j$, whence the root
bracket with $v_{C_j}=z\,u^{(j)}_C$. Each older member received the clique,
is present with probability $e^{-r\tau}$ and of age $\tau+w$, and so supplies
one factor $\mathcal A(z,\tau)$ apiece. The final identity in \eqref{eq:PhiR}
is the probability generating functional of the process, namely the
exponential of the integrated single-clique defect $1-H^{(j)}_R$; this is
Campbell's theorem in generating-function form. It is the point-process analogue of
$\langle x^{N}\rangle=e^{-\lambda(1-x)}$ for a Poisson count $N$ of mean
$\lambda=\gamma_j n_j\sigma$, generalised to the location-dependent
$x=H^{(j)}_R(z,\tau)$. Differentiating \eqref{eq:PhiR} in $\sigma$ gives the
Volterra form \eqref{eq:volterra-ode} of the main text.

\subsection{Cavity closure}

The cavity generating functions are the products of the channels that remain
when the incoming clique is deleted, the roles reversing across an edge
exactly as established in Appendix~\ref{sec:appendix-mp}. Hence
\begin{align}
    u_B(z,\sigma) &= \prod_{l=1}^{m}\big(H^{(l)}_C\big)^{n_l}
                     \prod_{l=1}^{m}D^{(l)}
                   = \mathcal A^{\varphi}\prod_{l=1}^{m}D^{(l)},
    \label{eq:uB}\\
    u^{(j)}_C(z,\sigma) &= H_B\,\big(H^{(j)}_C\big)^{n_j-1}\!
                     \prod_{l\ne j}\big(H^{(l)}_C\big)^{n_l}
                     \prod_{l=1}^{m}D^{(l)}
                   = H_B\,\frac{\mathcal A^{\varphi}}{\mathcal A^{\gamma_j}}
                     \prod_{l=1}^{m}D^{(l)},
    \label{eq:uC}\\
    u_R(z,\sigma) &= H_B\,\prod_{l=1}^{m}\big(H^{(l)}_C\big)^{n_l}
                     \prod_{l=1}^{m}D^{(l)}
                   = H_B\,\mathcal A^{\varphi}\prod_{l=1}^{m}D^{(l)},
    \label{eq:uR}
\end{align}
with $\varphi$ of Eq.~\eqref{eq:Gamma-def}; the received channel needs no
excess correction because a Poisson process is unchanged by the removal of
one of its points. Comparison with \eqref{eq:gf-product} gives the compact
statement
\begin{equation}
    G(z,\sigma) = z\,u_R(z,\sigma),
    \label{eq:G-is-zuR}
\end{equation}
and dividing the closures pairwise gives the algebraic links
\eqref{eq:cavity-links}. Setting $v_a=z\,u_a$ throughout (the reach along
a cavity channel, including the vertex reached) and inserting
\eqref{eq:PhiB-full}, \eqref{eq:PhiC} and \eqref{eq:PhiR} yields the system
\eqref{eq:vB}--\eqref{eq:vR} quoted in the main text, with
$G(z,\sigma)=v_R(z,\sigma)$.

\subsection{Reduction at $z=1$: recovering the giant component}
\label{app:z1-reduction}

The fixed point \eqref{eq:vB}--\eqref{eq:vR} contains the message passing of
Appendix~\ref{sec:appendix-mp} as its $z=1$ boundary. Since $G(z,\sigma)$
generates the finite-cluster probabilities, its value at $z=1$ is the total
probability that the component of an age-$\sigma$ vertex is finite; in the
locally tree-like limit this is precisely the probability $\mathcal V(\sigma)$ that
the vertex is not joined to the giant, evaluated under site percolation with
occupation $\phi$. We verify that the fixed point delivers exactly this.

Write $\mathcal V(\sigma)\equiv v_R(1,\sigma)$. The links
\eqref{eq:cavity-links} identify the remaining cavities at $z=1$ as the
excess probabilities of Appendix~\ref{sec:appendix-mp},
\begin{equation}
    v_B(1,\sigma)=\frac{\mathcal V(\sigma)}{H_B(1,\sigma)},
    \qquad
    v_{C_j}(1,\sigma)=\frac{\mathcal V(\sigma)}{\mathcal A(1,\sigma)^{\gamma_j}}.
    \label{eq:z1-cavities}
\end{equation}
It remains to check that each channel factor at $z=1$ coincides with its
counterpart in Appendix~\ref{sec:appendix-mp} after the site-percolation
substitution $\mathcal V\to(1-\phi)+\phi\,\mathcal V$ in every excess. For
the single-older-neighbour factor, the substitution applied to
Eq.~\eqref{eq:mp-att} gives
\begin{align}
    A(\sigma)
    &= 1-e^{-r\sigma}\Big(1-\big[(1-\phi)+\phi\,(\mathsf J\mathcal V)(\sigma)\big]\Big)
    \nonumber\\
    &= 1-\phi e^{-r\sigma}\big[1-(\mathsf J\mathcal V)(\sigma)\big]\nonumber\\
     &= \mathcal A(1,\sigma),
    \label{eq:z1-A}
\end{align}
the middle line being Eq.~\eqref{eq:calA} at $z=1$. The same one-line
computation applied to Eq.~\eqref{eq:mp-birth}, with the excess
$\mathcal V/B$ supplied by \eqref{eq:z1-cavities}, gives
$B(\sigma)=[1-\phi e^{-r\sigma}(1-\mathcal V/B)]^{\gamma_b}
=H_B(1,\sigma)$; and applied to the root bracket \eqref{eq:app-Rj}, with
the excess $\mathcal V/A^{\gamma_j}$, gives
\begin{align}
    H^{(j)}_R(1,\tau)
    =&\,\Big[1-\phi e^{-r\tau}\Big(1-\frac{\mathcal V(\tau)}{A(\tau)^{\gamma_j}}\Big)\Big]
    A(\tau)^{\gamma_j-1}\nonumber\\
    =&\,R_j(\tau)\,A(\tau)^{\gamma_j-1},
    \label{eq:z1-Hj}
\end{align}
the failure probability of one received clique, so that the exponent
\eqref{eq:PhiR} resums the received collection to
$\prod_j D^{(j)}(1,\sigma)=D(\sigma)$ of Eq.~\eqref{eq:app-D}.
Multiplying the channels, Eq.~\eqref{eq:vR} at $z=1$ reads
\begin{equation}
    \mathcal V(\sigma)
    = B(\sigma)\,\prod_j A(\sigma)^{\gamma_j n_j}\,D(\sigma),
    \label{eq:z1-fixedpoint}
\end{equation}
which is the fixed point \eqref{eq:app-factorise} of
Appendix~\ref{sec:appendix-mp} under site percolation; solved by the same
iteration on the half-line, $v_R(1,\sigma)$ is the site-percolated
$\mathcal V(\sigma)$. Below $\phi_c$ the only solution is
$\mathcal V\equiv1$ and the finite-cluster distribution is normalised,
$G(1,\sigma)=1$; above $\phi_c$ the deficit integrates to the giant
fraction, $\phi\int_0^\infty e^{-\sigma}[1-G(1,\sigma)]\,d\sigma=S$,
recovering Eq.~\eqref{eq:perc-S}. The giant component of
Sec.~\ref{sec:giant} is therefore the $z=1$ corollary of the
component-size fixed point, and the agreement of the two independently
constructed recursions (the scalar reach probability of Appendix~%
\ref{sec:appendix-mp} and the generating-function closure of the present
appendix) is an internal consistency check on both.

\subsection{Linearisation of the fixed point}

The system \eqref{eq:vB}--\eqref{eq:vR} is a \emph{vector} fixed point
$\mathbf v=(v_B,v_{C_1},\dots,v_{C_m},v_R)$ of the form
$v_a=z\,\Psi_a(\mathbf v)$ on $m+2$ components (the inert $v_{C_j}$ with
$n_j=0$ dropped), with
\begin{align}
    \Psi_B &= \mathcal A^{\varphi}\,\textstyle\prod_{l}D^{(l)},
    \nonumber\\
    \Psi_{C_j} &= H_B\,\mathcal A^{\varphi-\gamma_j}\,
    \textstyle\prod_{l}D^{(l)},
    \label{eq:Psi-defs}\\
    \Psi_R &= H_B\,\mathcal A^{\varphi}\,\textstyle\prod_{l}D^{(l)}.
    \nonumber
\end{align}
By the product rule, the functional derivatives of the $\Psi_a$ assemble
from four elementary kernels,
\begin{align}
    \frac{\delta H_B(\sigma)}{\delta v_B(\sigma')}
    =&\, \gamma_b\,\phi e^{-r\sigma}
    \big[1-\phi e^{-r\sigma}\big(1-v_B(\sigma)\big)\big]^{\gamma_b-1}
    \delta(\sigma-\sigma'),
    \label{eq:fjac-local}\\
    \frac{\delta\mathcal A(\sigma)}{\delta v_R(\sigma')}
    =&\, \phi e^{-r\sigma}\,e^{-(\sigma'-\sigma)}\,\Theta(\sigma'-\sigma),
    \label{eq:fjac-B}\\
    \frac{\delta\ln D^{(j)}(\sigma)}{\delta v_{C_j}(\sigma')}
    =&\, \gamma_j n_j\,\phi e^{-r\sigma'}\,
    \mathcal A(z,\sigma')^{\gamma_j-1}\,\Theta(\sigma-\sigma'),
    \label{eq:fjac-R}\\
    \frac{\delta\ln D^{(j)}(\sigma)}{\delta v_R(\sigma')}
    =&\, \gamma_j n_j(\gamma_j-1)\,\phi\!\!\int_0^{\min(\sigma,\sigma')}
    \!\!\!\! e^{-r\tau}
    \big[1-\phi e^{-r\tau}\big(1-v_{C_j}(\tau)\big)\big]\nonumber\\
    &\times\,
    \mathcal A(z,\tau)^{\gamma_j-2}\,e^{-(\sigma'-\tau)}\,d\tau,
    \label{eq:fjac-2sided}
\end{align}
\end{widetext}
with $\Theta$ the step function: a local multiplication kernel, the forward
exponential kernel of $\mathsf J$, the backward Volterra kernel, and, for
$\gamma_j\ge2$ only, their composition, supported on \emph{both} sides of
the diagonal. These kernels carry the constructive content of the remainder
of the appendix: contracted against the profiles
$\nu_a=\partial_z v_a|_{z=1}$ they produce the moment equations
\eqref{eq:nuR-sum}--\eqref{eq:R-def} of the main text, and the $z$-expansion
of the next section integrates lower-order profiles against the same
kernels, order by order.

\paragraph*{Relation to Lagrange--Good inversion.}
For finitely many types the fixed point $v_a=z\,\Psi_a(\mathbf v)$ is algebraic, and the Lagrange--Good formula~\cite{Good1960} resums the tree expansion it generates into a Jacobian determinant. This is the typical route to the
closed-form component distributions of the configuration
model~\cite{Newman2007,PhysRevE.95.052303}. Functional extensions of the formula are known, to integral equations~\cite{deBruijn1983} and to an uncountable ``colour palette''~\cite{Jansen2021}, the colour here being the pair of channel and age: the Jacobian becomes a Fredholm determinant of
infinitesimal kernels, precisely the operators~\eqref{eq:fjac-local}--\eqref{eq:fjac-2sided}. For the variable-limit forward--backward kernels above, however, this resummation is only formal, its term-by-term content being exactly the sum over age-labelled trees that the recursion~\eqref{eq:hierarchy} evaluates by quadrature. Lagrange--Good inversion therefore does not fail but degenerates into the constructive hierarchy; what is lost is only the closed-form resummation, and with it the age-locality that permitted it.

\subsection{Moments, the hierarchy, and the base order}

The moment equations quoted in the main text follow by differentiating
\eqref{eq:vR} at $z=1$. Below threshold every cavity equals unity there,
every bracket and every $H^{(j)}_R$ collapses to $1$, and logarithmic
differentiation of the product gives Eq.~\eqref{eq:nuR-sum} with the channel
derivatives \eqref{eq:BC-def} and \eqref{eq:R-def}: $\hat B=\partial_z H_B|_1$,
$\hat C_j=\partial_z\mathcal A^{\gamma_j}|_1$ per created clique, and
$\hat R_j=\partial_z D^{(j)}|_1$, the last obtained by differentiating
\eqref{eq:PhiR} under the integral, where $\partial_z H^{(j)}_R|_1=\phi e^{-r\tau}
\nu_{C_j}(\tau)+(\gamma_j-1)\phi e^{-r\tau}(\mathsf J\nu_R)(\tau)$.
Differentiating the links \eqref{eq:cavity-links} at $z=1$ gives the closure
$\nu_B=\nu_R-\hat B$ and $\nu_{C_j}=\nu_R-\hat C_j$; in particular
$\hat B=\gamma_b\phi e^{-r\sigma}(\nu_R-\hat B)$, so
$\hat B=\gamma_b\phi e^{-r\sigma}\nu_R/(1+\gamma_b\phi e^{-r\sigma})$, as quoted
in the main text. Every higher moment is likewise a linear integral
equation, obtained one order at a time.

The coefficient hierarchy \eqref{eq:hierarchy} is explicit because the
$\Psi_a$ carry no explicit factor of $z$: the coefficient of $z^{s-1}$ in
$\Psi_a(\mathbf v(z,\cdot);\sigma)$ involves only the profiles $c^{(p)}_a$
with $p\le s-1$, already known on the whole half-line, so each order is a
quadrature of lower-order profiles against the fixed kernels
\eqref{eq:fjac-B}--\eqref{eq:fjac-2sided} and no equation need be inverted.
The base order quoted in Eq.~\eqref{eq:pi1} is
$\theta_1(\sigma)=\Psi_R(\mathbf 0;\sigma)$: with $\mathbf v=\mathbf 0$ one has
$\mathcal A=1-\phi e^{-r\sigma}$ and $H^{(j)}_R=(1-\phi e^{-r\tau})^{\gamma_j}$,
whence the exponents $\gamma_b+\varphi$ and the survival integrals of
\eqref{eq:pi1}, whose exponent is elementary,
\begin{equation}
    \int_0^\sigma\!\Big[1-\big(1-\phi e^{-r\tau}\big)^{\gamma}\Big]d\tau
    = \sum_{q=1}^{\gamma}\binom{\gamma}{q}
    \frac{(-1)^{q+1}\phi^{q}}{q\,r}\big(1-e^{-q r\sigma}\big).
    \label{eq:pi1-integral}
\end{equation}

Nothing in the construction, finally, ties the injection to a single vector:
were incoming vertices to draw their joint degrees independently from a
distribution $q(\alpha'_1,\dots,\alpha'_m)$, the created factor of
\eqref{eq:gf-product} would become the $q$-mixture of
$\prod_j[H^{(j)}_C]^{n_j}$, the received rates would be replaced by their
means $\gamma_j\langle n_j\rangle_q$ with the root bracket of
\eqref{eq:Hj} carrying the size-biased law
$n_j\,q(\bm\alpha')/\langle n_j\rangle_q$, and the hierarchy
\eqref{eq:hierarchy} would go through unchanged.

\begin{acknowledgments}
The authors would like to thank an anonymous reviewer for their very helpful comments and suggestions to an earlier version of the manuscript.
\end{acknowledgments}

\bibliography{ref}
\end{document}